\documentclass[%
twocolumn,
superscriptaddress,
 amsmath,amssymb,
 aps,
prl,
noeprint,
noinfoline
]{revtex4-2}

\usepackage{graphicx}% Include figure files
\usepackage{dcolumn}% Align table columns on decimal point
\usepackage{bm}% bold math
\usepackage{color}

\begin{document}

\preprint{APS/123-QED}

\title{On-Demand Magnon Resonance Isolation in Cavity Magnonics}

\author{Amin Pishehvar}
\affiliation{ 
    Department of Electrical and Computer Engineering, Northeastern University, Boston, MA 02115, USA
}

\author{Zhaoyou Wang}
\affiliation{ 
    Pritzker School of Molecular Engineering, University of Chicago, Chicago, IL 60637, USA
}

\author{Yujie Zhu}
\affiliation{ 
    Department of Materials Science and Engineering, University of Wisconsin - Madison, Madison, WI 53706, USA
}

\author{Yu Jiang}
\affiliation{ 
    Department of Electrical and Computer Engineering, Northeastern University, Boston, MA 02115, USA
}

\author{Zixin Yan}
\affiliation{ 
    Department of Electrical and Computer Engineering, Northeastern University, Boston, MA 02115, USA
}

\author{Fangxin Li}
\affiliation{ 
    Pritzker School of Molecular Engineering, University of Chicago, Chicago, IL 60637, USA
}

\author{Josep M. Jornet}
\affiliation{ 
    Department of Electrical and Computer Engineering, Northeastern University, Boston, MA 02115, USA
}

\author{Jia-Mian Hu}
\affiliation{ 
    Department of Materials Science and Engineering, University of Wisconsin - Madison, Madison, WI 53706, USA
}

\author{Liang Jiang}
\affiliation{ 
    Pritzker School of Molecular Engineering, University of Chicago, Chicago, IL 60637, USA
}

\author{Xufeng Zhang}
\email{xu.zhang@northeastern.edu}
\affiliation{ 
    Department of Electrical and Computer Engineering, Northeastern University, Boston, MA 02115, USA
}
\affiliation{ 
    Department of Physics, Northeastern University, Boston, MA 02115, USA
}

\date{\today}% It is always \today, today,
             %  but any date may be explicitly specified

\begin{abstract}
Cavity magnonics is a promising field focusing the interaction between spin waves (magnons) and other types of signals. In cavity magnonics, the function of isolating magnons from the cavity to allow signal storage and processing fully in the magnonic domain is highly desired, but its realization is often hindered by the lack of necessary tunability on the interaction. This work shows that by utilizing the collective mode of two YIG spheres and adopting Floquet engineering, magnonic signals can be switched on-demand to a magnon dark mode that is protected from the environment, enabling a variety of manipulation over the magnon dynamics. Our demonstration can be scaled up to systems with an array of magnonic resonators, paving the way for large-scale programmable hybrid magnonic circuits.
\end{abstract}

%\keywords{Suggested keywords}

\maketitle

%\tableofcontents

%%%%%%%%%%%%%%%%%%%%%%%%%%%%%%%%%%%%%%%%%%%%%%%%%%%%%%%%%%%%%%%%
%%%%%%%%%%%%%%%%%%%%%%%%%%%%%%%%%%%%%%%%%%%%%%%%%%%%%%%%%%%%%%%%
%%%
%\section{Introduction} 
%%%
%%%%%%%%%%%%%%%%%%%%%%%%%%%%%%%%%%%%%%%%%%%%%%%%%%%%%%%%%%%%%%%%
%%%%%%%%%%%%%%%%%%%%%%%%%%%%%%%%%%%%%%%%%%%%%%%%%%%%%%%%%%%%%%%%

Dark mode \cite{Benisty2009Apr} in electromagnetic devices is a phenomenon where a mode is isolated from the environment due to destructive interference or suppressed coupling. These modes often exhibit extended lifetimes because of reduced radiation emission into the environment, leading to important applications across a wide variety of platforms. For instance, in integrated photonics, dark modes are implemented in metasurfaces to achieve supercavity lasing \cite{Rybin2017Dec,Rybin2017Jan,Kodigala2017Jan}; in optomechanics, dark mode allows phonon-mediated quantum coupling between two optical resonators without requiring the mechanical resonator cooled down to its quantum mechanical ground state \cite{Dong2012Nov,Wang2012Apr,Wang2012Oct,Lai2020Aug_PRA}; while for superconducting qubits or cold atoms, dark mode is used to protect delicate quantum states \cite{Zanner2022May,White2019Jun}.

In the emerging field of cavity magnonics \cite{Rameshti_PhysRep_2022,Harder_SSC_2018,Lachance_APE_2019,Bhoi_SSP_2020,YiLi_JAP_2020,Awschalom2021Feb,Zhang2023Sep_MTE}, dark mode also finds important applications. Cavity magnonics studies the interaction between magnons and microwave photons in hybrid devices \cite{Zhang2014Oct,Tabuchi2014Aug,Goryachev2014Nov,Bai2015Jun,Li2019Sep_PRL,Hou2019Sep,Wang2022Dec}, with promising potential in coherent and quantum information processing \cite{Tabuchi2015Jul,Lachance-Quirion2020Jan,Lachance-Quirion2017Jul,Wolski2020Sep,Xu2023May_PRL}. Thanks to the large spin density in magnonic resonators, the magnon-photon coupling strength can exceed their individual dissipation, bringing the system into the strong coupling regime, where information is exchanged between magnon and photon modes multiple times before decaying to below the noise level \cite{Zhang2014Oct,Zhang2015Nov_gradientmemory,Wolz2020Jan,Song2023Sep}, enabling critical signal processing functionalities \cite{Xu2021May_PRL}. When multiple magnonic resonators couple with a single microwave cavity, collective magnon dark modes can form, which isolate magnons from the cavity \cite{Zhang2015Nov_gradientmemory,Li2022Jan_PRL}, allowing the information to be stored in the magnonic domain and fully processed using magnonic approaches.

To harness the full potential of magnon dark modes, on-demand dark-bright mode conversion is required. However, such mode conversion is a fundamental challenge on all physical platforms, considering the isolated nature of the dark modes. It is worth noting that such on-demand conversion is fundamentally different -- and thus should be differentiated -- from the fixed dark-bright mode coupling which is relatively easy to achieve \cite{Cao2013Sep_APL,Zhang2016Feb_SciAdv,Rodriguez2011Dec_PRX,Panaro2014Apr_ACSPhoton,Meng2023Nov_OE}. Thus far, on-demand dark-bright mode conversion has only been demonstrated on very few systems, such as integrated photonics \cite{Zhang2019Jan_NP} and optomechanics \cite{Lake2020May_NC}. In this work, we show that, for the first time, on-demand dark-bright mode conversion can be achieved on a magnetic platform. By applying Floquet engineering \cite{Xu2020Dec} to a multimode magnonic system, the magnon dark and bright modes can be coupled with a coupling strength determined by the amplitude and phase of a driving signal. This capability enables on-demand isolation of the giant spin ensemble from the microwave cavity, paving the way for advanced magnonic signal processing.

\begin{figure}[bt]
    \centering
    \includegraphics[width=0.95\linewidth]{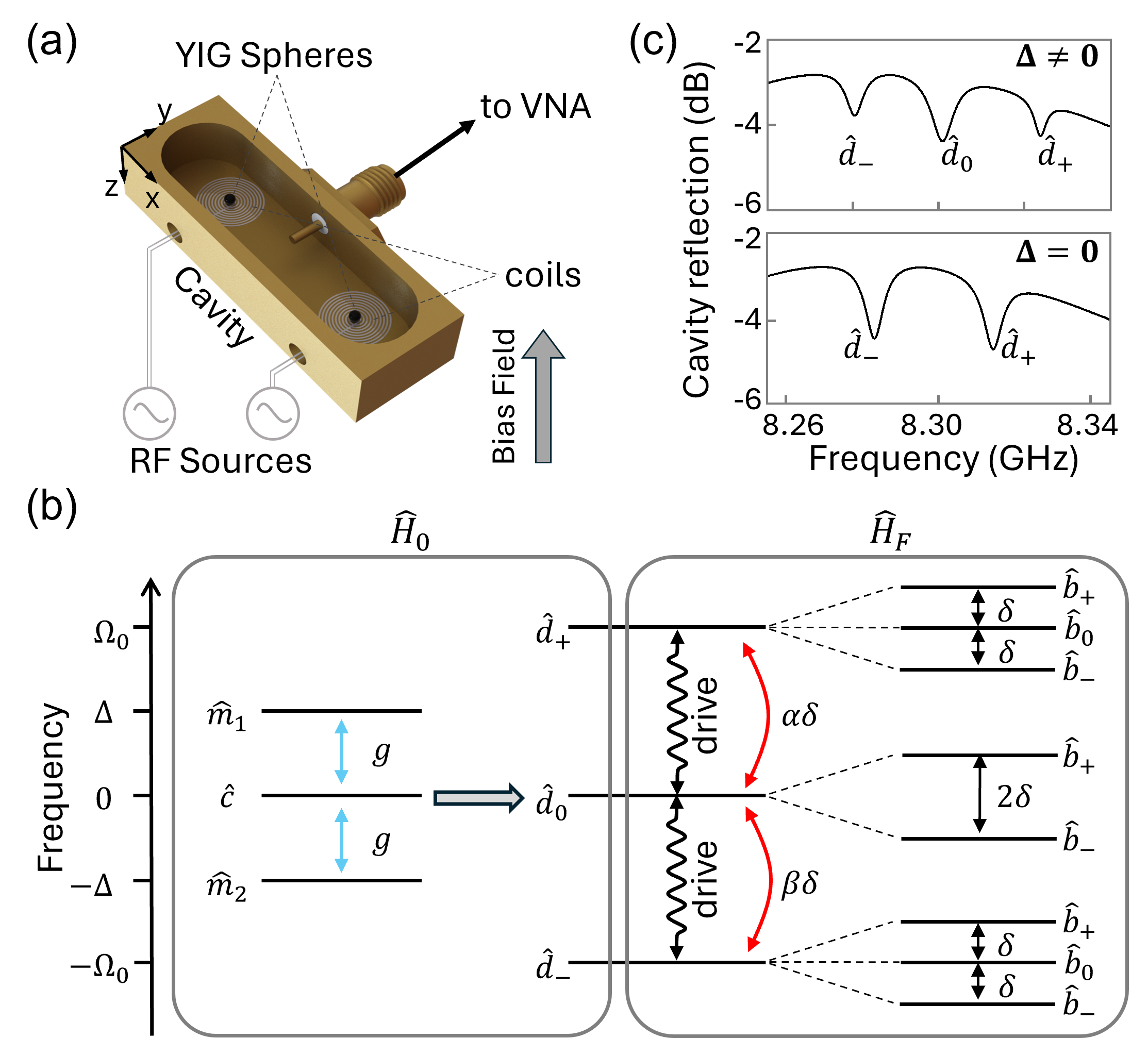}
    \caption{(a) Device schematics. Two YIG spheres are placed at the center of two flat coils inside a copper cavity. The coils are parallel to the bias magnetic field to applying the Floquet drive. The coaxial probe is for microwave excitation and readout. (b) Energy level diagram. $\hat{m}_1 (\hat{m}_2)$, first (second) magnon mode; $\hat{c}$, cavity mode; $g$, magnon-photon coupling strength; $\hat{d}_\pm$, upper (lower) hybrid mode; $\hat{d}_0$, center hybrid mode; $\alpha_\pm\delta$, energy level difference of $\hat{d}_0$ with $\hat{d}_\pm$; $\hat{b}_\pm$, upper (lower) Floquet hybrid mode; $\hat{b}_0$, center Floquet hybrid mode; $\delta$, energy level difference between two Floquet hybrid modes. (c) Reflection spectra of the microwave cavity. Top: three modes ($\hat{d}_0$ and $\hat{d}_\pm$) from the hybridization of one cavity mode and two magnon modes from the two YIG spheres; bottom: the center mode ($\hat{d}_0$) becomes a dark mode and disappears when $\Delta=\omega_m-\omega_c=0$.}
    \label{fig1}
\end{figure}

%%%%%%%%%%%%%%%%%%%%%%%%%%%%%%%%%%%%%%%%%%%%%%%%%%%%%%%%%%%%%%%%
%%%%%%%%%%%%%%%%%%%%%%%%%%%%%%%%%%%%%%%%%%%%%%%%%%%%%%%%%%%%%%%%
%%%
%\section{Device and concept} 
%%%
%%%%%%%%%%%%%%%%%%%%%%%%%%%%%%%%%%%%%%%%%%%%%%%%%%%%%%%%%%%%%%%%
%%%%%%%%%%%%%%%%%%%%%%%%%%%%%%%%%%%%%%%%%%%%%%%%%%%%%%%%%%%%%%%%

Our device comprises a three-dimensional (3D) microwave cavity supporting a $\mathrm{TE}_{101}$ mode at $\omega_c=2\pi\times8.3$ GHz with a quality factor of 2500 [Fig. \ref{fig1}(a)]. The cavity is probed by a coaxial probe via reflection measurement (S$_{11}$) using a vector network analyzer (VNA). Two identical yttrium iron garnet (YIG) spheres (diameter: 0.4 mm) are positioned at the bottom of the cavity, each supporting a Kittel mode at frequency $\omega_n=\gamma H_n$, where $H_n$ is the bias magnetic field for the $n$-th ($n=1,2$) YIG sphere provided by a permanent magnet, and $\gamma=2\pi\times28$ GHz/T is the gyromagnetic ratio. To introduce Floquet drives to the magnon modes, a 10-turn flat coil is placed inside the cavity underneath each YIG sphere, with the coil axis parallel to the bias field direction ($z$ direction).

This multimode Floquet cavity electromagnonic system can be described by the Hamiltonian $\hat{H}(t)=\hat{H}_0+\hat{H}_F(t)$. The first term represents the conventional cavity electromagnonic system, which in the rotating frame of the cavity mode can be represented as
\begin{equation}
\hat{H}_0=\hbar\sum_{n=1,2}\left[ \Delta_n\hat{m}_n^\dag\hat{m}_n+g_n(\hat{c}^\dag \hat{m}_n + \hat{c} \hat{m}_n^\dag)\right],
\label{Eq:H0}
\end{equation}
\noindent where %$n=1,2$ is the YIG sphere index, 
$\Delta_n=\omega_n-\omega_c$ is the frequency detuning of the $n$-th magnon mode, $g_n$ is the coupling strength between the $n$-th magnon mode and the cavity mode, $\hat{c}^\dag$ and $\hat{c}$ ($\hat{m}_n^\dag$ and $\hat{m}_n$) are the creation and annihiliation operators for the cavity ($n$-th magnon) mode, respectively. Unless explicitly mentioned, we assume $g_1=g_2=g$ and $\Delta_1=-\Delta_2=\Delta>0$ throughout our analysis. %Accordingly, the magnon mode in sphere 1 (2) has a higher (lower) frequency than the cavity resonance. 
The second term $\hat{H}_F(t)$ represent the Floquet interaction
\begin{equation}
    \hat{H}_F(t) = \hbar \sum_{n={1,2}}{\epsilon}_n \cos({\Omega}_F t+\phi_n) {\hat{m}_n^\dag} \hat{m}_n,
    \label{Eq:HF}
\end{equation}
where $\epsilon_n$ and $\phi_n$ are the amplitude and phase of the Floquet drive applied to the $n$-th YIG sphere, respectively. $\Omega_F$ is the frequency of the Floquet drive, which are assumed to be identical for both YIG spheres for simplicity.

The system can be illustrated by the energy level diagram in Fig.\,\ref{fig1}(b). The magnon modes (Kittel modes) of the two YIG spheres couple simultaneously to the same cavity mode, rending the two spheres an effective giant spin ensemble. In the absence of the Floquet drive, the two magnon modes form two ensemble modes $\hat{m}_\mathrm{B}=(\hat{m}_1+\hat{m}_2)/\sqrt{2}$ and $\hat{m}_\mathrm{D}=(\hat{m}_1-\hat{m}_2)/\sqrt{2}$. When both magnon modes are on resonance with the cavity ($\Delta=0$), $\hat{m}_\mathrm{D}$ does not couple with the cavity mode $\hat{c}$ (thus referred to as the magnon dark mode \cite{Zhang2015Nov_gradientmemory}), leading to the final dark mode $\hat{d}_0=\hat{m}_\mathrm{D}$, which cannot be observed in the cavity reflection spectrum; while
$\hat{m}_\mathrm{B}$ is coupled to the cavity mode $\hat{c}$ (accordingly referred to as the magnon bright mode), forming two new normal modes $\hat{d}_\pm$ at $\omega_c\pm\Omega_0$ [Fig.\,\ref{fig1} (c)]. Under this condition, the coupling of a single magnon mode with the cavity mode can be extracted from the splitting $g/2\pi = \Omega_0/2\sqrt{2}\pi = 12.6$ MHz. When the magnon modes are detuned ($\Delta \neq 0$), the magnon dark mode becomes ``less dark'' and gradually shows up in the cavity reflection spectrum.

When a Floquet drive with a frequency $\Omega_\mathrm{F}=\Omega_0$ is applied, it facilitates the interaction between the hybrid modes, as indicated by the Floquet Hamiltonian in the rotating frame of $\hat{H}_0$
\begin{equation}
    \hat{H}_F=\delta\hat{d}_0^\dag(\alpha \hat{d}_+ + \beta\hat{d}_-)=\delta \hat{d}_0^\dag \hat{d}_\mathrm{B},
    \label{Eq:HF_RoratingFrame}
\end{equation}
where $\alpha$ and $\beta$ are the mixing coefficients \cite{SM}. Here $\hat{d}_\mathrm{B}= \alpha \hat{d}_+ + \beta \hat{d}_-$ originates from the superposition of $\hat{d}_+$ and $\hat{d}_-$, which is referred to as the Floquet bright mode because it couples (assisted by the Floquet drive) to the center mode $\hat{d}_0$ with a nonzero coupling strength $\delta$, forming two new hybrid modes $\hat{b}_\pm$. In contrast, the orthogonal superposition mode $\hat{d}_\mathrm{D}= \beta^* \hat{d}_+ - \alpha^* \hat{d}_-$ does not couple with the $\hat{d}_0$ mode, and thus is referred to as the Floquet dark mode. The Floquet bright (dark) mode can also be viewed as the result of constructive (destructive) interference between the Floquet couplings from the $\hat{d}_\pm$ to the $\hat{d}_0$ mode. As a result of these interactions, mode $\hat{d}_0$ splits into two levels associated with $\hat{b}_\pm$ which are separated by $2\delta$ (assuming $\alpha=\beta=1$ for simplicity but without losing generalicity), while the third level $\hat{b}_0=\hat{d}_\mathrm{D}$ disappears from the spectrum because of the cancelled coupling. Similarly, $\hat{d}_\pm$ also each splits into two levels associated with $\hat{b}_\pm$ with a separation of $2\delta$. However, the coupling of $\hat{d}_0$ with $\hat{d}_+$ (or $\hat{d}_-$) does not cancel for $\hat{d}_+$ mode (or $\hat{d}_-$), and therefore it can still be observed [$\hat{b}_0$ modes in Fig.\,\ref{fig1}(c)].

\begin{figure}[bt]
    \includegraphics[width=0.95\linewidth]{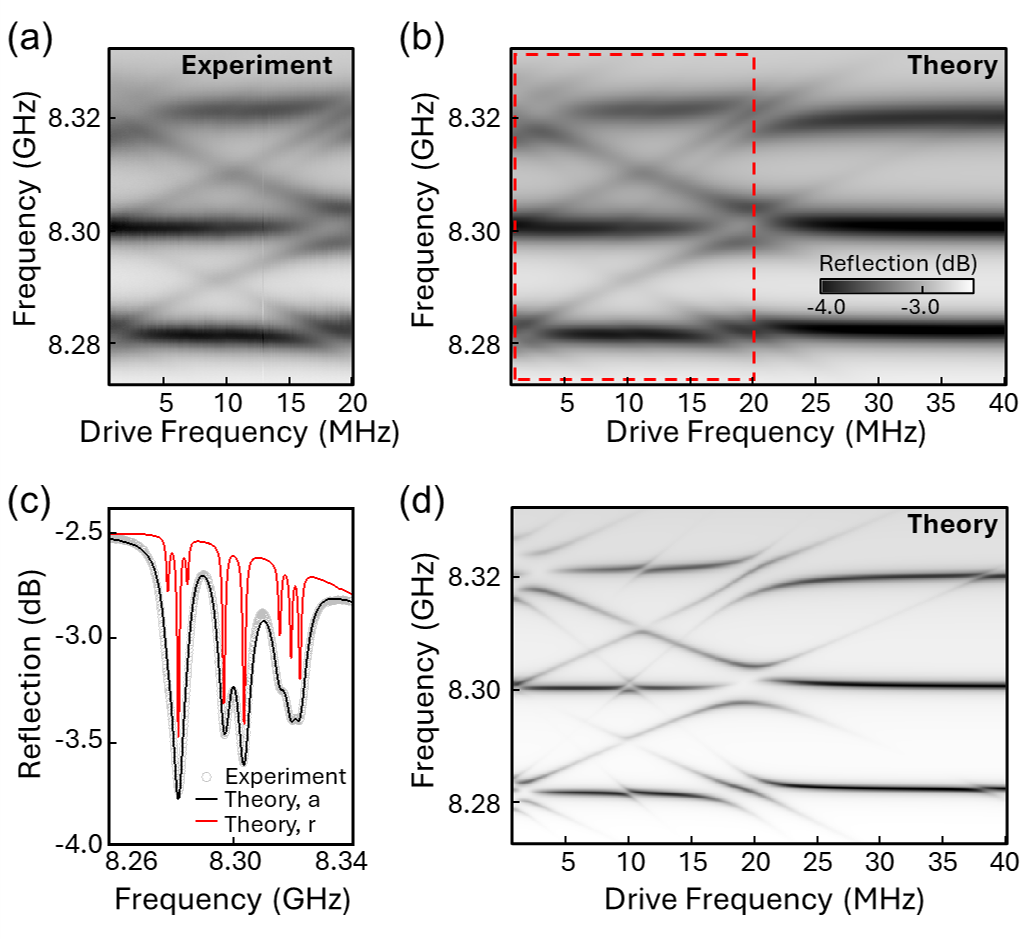}
    \caption{(a),(b) Measured and calculated cavity reflection spectra as a function of driving frequency when both YIG spheres are driven, respectively. (c) Line plot of the measured cavity reflection (circles) and the numerical fitting (black line), together with calculation results with reduced linewidth (red). (d) Calculated cavity reflection spectra using reduced linewidth.}
    \label{fig2}
\end{figure}

\begin{figure}[tb]
    \centering
    \includegraphics[width=0.95\linewidth]{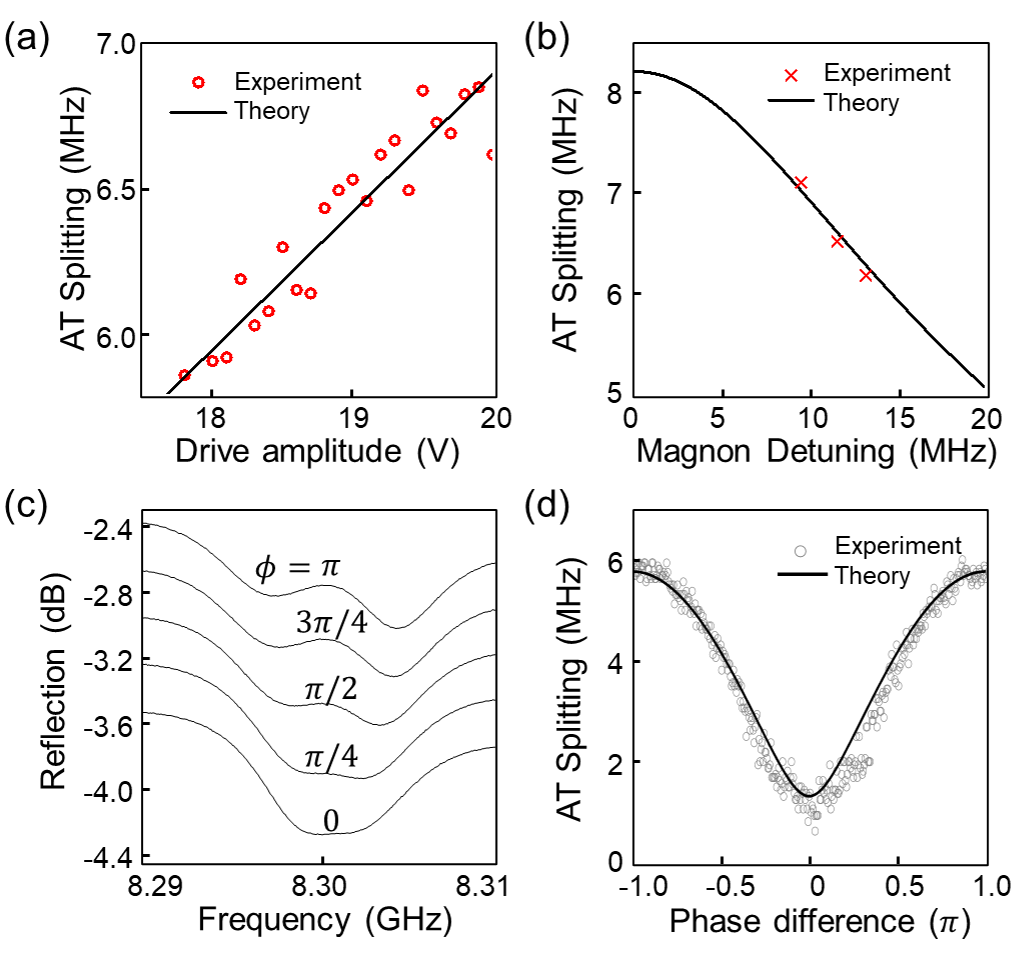}
    \caption{(a)-(b) AT splitting as a function of the drive amplitude (here $\epsilon_1=\epsilon_2$) and magnon detuning, respectively. (c) Measured cavity reflection spectra at different values of the phase difference $\phi$ between the two Floquet drives. (d) Extracted AT splitting as a function of the phase difference $\phi$.}
    \label{fig3}
\end{figure}

Figure\,\ref{fig2} plots the measured cavity reflection spectra from a device with a nonzero magnon detuning ($\Delta > 0$). Three hybrid modes are observed at 8.32103 GHz ($\hat{d}_+$), 8.30365 GHz ($\hat{d}_0$), and 8.28434 GHz ($\hat{d}_-$). When a Floquet drive is applied to both YIG spheres, Autler-Townes (AT) splittings show up on all the three modes. As the drive frequency is tuned to match the level separation ($\Omega_\mathrm{F}=\Omega_0=2\pi\times 18.55$ MHz), the Floquet bright mode $\hat{d}_\mathbf{B}$ is fully hybridized with the center mode $\hat{d}_0$. Here the center mode $\hat{d}_0$ splits into two hybrid modes ($\hat{b}_\pm$), and no modes can be observed at the original frequency of the $\hat{d}_0$ mode, because the Floquet dark mode $\hat{d}_\mathbf{D}$ completely decouples with the $\hat{d}_0$ mode. Once the drive frequency shifts away from $\Omega_0$, the two hybrids are further apart, and a center mode gradually shows up because the Floquet dark mode is no longer completely ``dark''. 

All the experimental observation in Fig.\,\ref{fig2}(a) agrees very well with our numerical modeling (see Supplemental Materials \cite{SM} for details), as shown by the calculated spectra in Fig.\,\ref{fig2}(b). The parameters used in the calculation is obtained through numerical fitting (details see Ref. \cite{SM}), which shows excellent agreement with the measurement results [Fig.\,\ref{fig2}(c)]. Our numerical fitting reveals that the Floquet drive on Sphere 2 is $40\%$ weaker than on Sphere 1, indicating that Sphere 2 is mounted at a larger distance from the surface of the driving coil. This can also explain the dissipation rates ($\kappa_2/2\pi = 3.1$ MHz $<$ $\kappa_1/2\pi=6.3$ MHz) obtained via the same numerical fitting: Sphere 1 suffers from higher losses due to closer proximity to the metallic coil. The spectra calculated using the same parameter set except for a five times reduction for all dissipation rates are also plotted [Fig.\,\ref{fig2}(d)], revealing more spectral details for the multi-mode Floquet coupling. For example, the three modes split from $\hat{d}_\pm$ become clearly visible. In addition, it also shows that the Floquet drive creates several high-order sidebands for each mode, which induce more anti-crossing features when crossing the other modes. Direct coupling between the $\hat{d}_\pm$ modes without involving $\hat{d}_0$ mode at around a drive frequency of $\Omega_F/2\pi=38$ MHz is also observed but with much smaller splittings.

\begin{figure*}[t]
    \centering
    \includegraphics[width=0.98\linewidth]{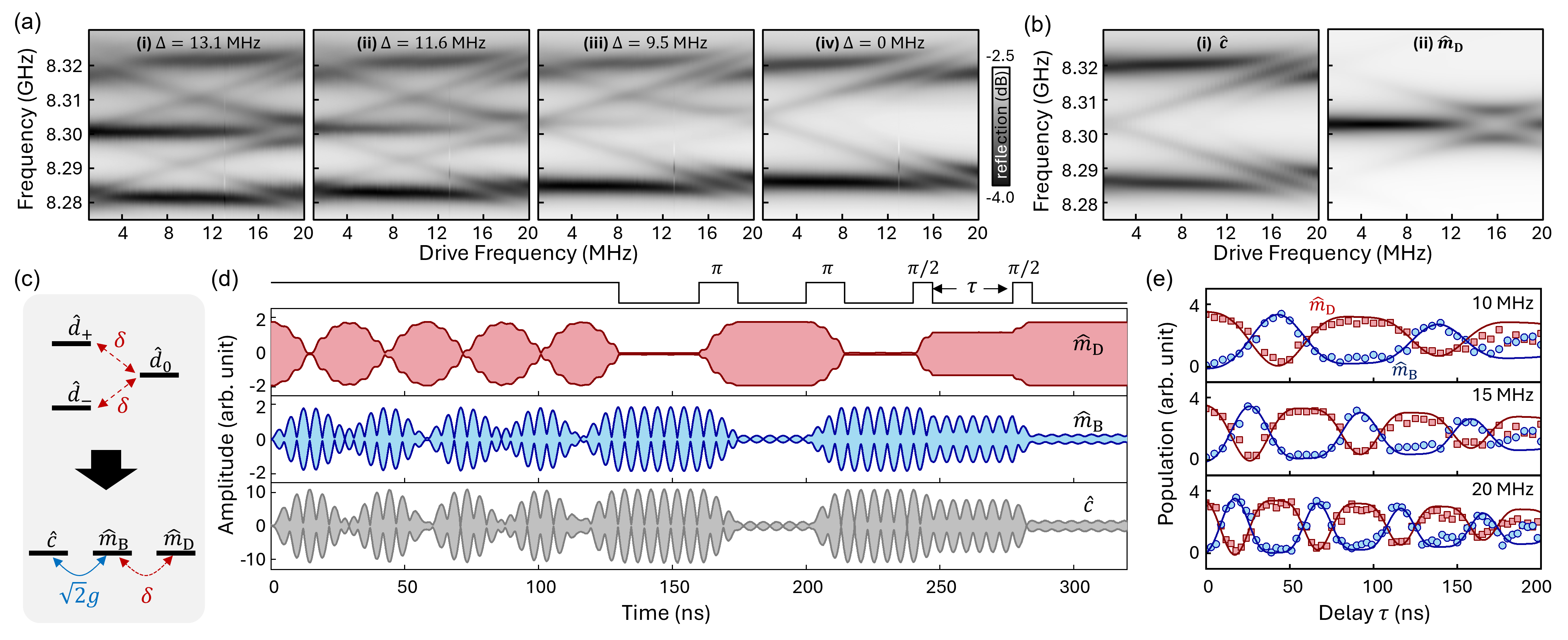}
    \caption{(a) Measured cavity photon ($\hat{c}$ reflection spectra at different magnon detunings: $\Delta = 13.2, 11.6, 9.5, 0$ MHz, respectively. (b) Calculated spectra of the cavity photon ($\hat{c}$) and magnon dark mode ($\hat{m}_\mathrm{D}$) using the analytical model, respectively. (c) Effective energy diagram of the system. (d) Dynamical phase-field simulation results of the pulse response of the actual system in (a) with $\Delta=0$ and $\Omega_F=\Omega_0$. The system is initially set to $\hat{m}_\mathrm{D}$ state at time 0. The top square waves show the Floquet drive pulses. (e) Magnon bright and dark mode amplitude obtained from the dynamical phase-field simulation at different drive detuning: $(\Omega_F-\Omega_0)/2\pi=10, 15, 20$ MHz, respectively. The signal is extracted after the application of a Ramsey pulse sequence and plotted as a function of the delay between the two $\pi/2$ pulse sequence. Red squares: dark mode from simulation; blue circles: bright mode from simulation. Solid lines are from analytical calculation.}
    \label{fig4}
\end{figure*}

Under the dual-drive condition, the Floquet coupling strength obtained from our model is \cite{SM}
\begin{equation}
\delta = \frac{g\sqrt{(\epsilon_1^2+\epsilon_2^2)(g^2+\Delta^2)-2\epsilon_1\epsilon_2 g^2\cos(\phi)}}{2(2g^2+\Delta^2)},    
\label{Eq:delta}
\end{equation}
where $\epsilon_n$ is the driving strength for the $n$-th sphere, $\phi = \phi_1-\phi_2$ is the phase difference between the two Floquet drives, respectively. The AT splitting between the $\hat{b}_\pm$ modes equals $2\delta$. According to Eq.\,\ref{Eq:delta}, the AT splitting exhibits a linear dependence on the driving amplitude $\epsilon$. This is confirmed by the extracted AT splittings at different driving amplitudes, as shown in Fig.\,\ref{fig3}(a). When a maximum driving amplitude (4 dBm) is applied, the AT splitting reaches 6.8 MHz. This value exceeds the dissipation rates of the the interacting modes ($\kappa_0/2\pi = 2.6$ MHz, $\kappa_+/2\pi = 3.4$ MHz, $\kappa_-/2\pi = 2.55$ MHz), indicating that the Floquet-drive-induced coupling has reached the strong coupling regime. The 3-mode Floquet interaction also depends on the detuning ($\Delta$) of the magnon modes. As the detuning reduces, the two hybrid modes $\hat{d}_\pm$ becomes closer in frequency, and accordingly it requires smaller drive frequencies. In the meantime, a larger AT splitting is achieved [crosses in Fig.\,\ref{fig3}(b)], which agrees with the theoretical prediction obtained using Eq.\,(\ref{Eq:delta}) [solid line in Fig.\,\ref{fig3}(b)].

In our experiments, the measured AT splitting shows a strong dependence on the relative phase $\phi$ of the two Floquet drives for the two YIG spheres [Fig.\,\ref{fig3} (c)]. When the two drives are out-of-phase ($\pm\pi$), $\hat{d}_0$ exhibits maximum AT splitting, corresponding to the constructive interference of the coupling with the two modes $\hat{d}_\pm$,where the coupling of $\hat{d}_0$ with $\hat{d}_\pm$ are in phase and add up to a stronger coupling. As $\phi$ decreases, the AT splitting becomes smaller, which reaches to a minimum when the two drives becomes in phase ($\phi=0$). In this case, the coupling of $\hat{d}_0$ with $\hat{d}_\pm$ are out of phase and cancel each other, leading to diminished coupling. Such phase dependence can be conveniently explained by our model in Eq.\,(\ref{Eq:delta}). The calculated splitting is plotted as a function of $\phi$ [solid line in Fig.\,\ref{fig3}(d)], which shows good agreement with the experimental results. The nonzero minimum splitting at $\phi = 0$ can be attributed to the nonzero detuning of the magnon modes.

Based on the dual-mode Floquet coupling demonstrated above, controlled dark-bright mode switching can be achieved. In the spectra shown in Fig.\,\ref{fig2}(a), the detuning of the two magnon modes is not zero, and thus the center mode $\hat{d}_0$ is not a pure magnon dark mode $\hat{m}_\mathrm{D}$ and thus can be observed. When the magnon detuning decreases, the center mode gradually disappears, as shown in Fig.\,\ref{fig4}(a). As $\Delta$ reaches zero, the center mode $\hat{d}_0$ becomes the magnon dark mode $\hat{m}_\mathrm{D}$ and thus completely disappear from the spectrum [Fig.\,\ref{fig4}(a)(iv)]. However, its interaction with $\hat{d}_\pm$ still exists, as indicated by the AT splitting in the lower and upper levels. Although the AT splitting in the center mode cannot be directly measured, it can be revealed by our numerical modeling, as shown in Fig.\,\ref{fig4}(b)(ii), where a splitting is clearly visible in the center mode of the calculated spectra for the magnon dark mode $\hat{m}_\mathrm{D}$. The validity of our model is verified by the calculated cavity reflection spectra [Fig.\,\ref{fig4}(b)(i)] which reproduced the experimental results in Fig.\,\ref{fig4}(a)(iv) with high accuracy. 

In fact, under the zero detuning condition $\Delta = 0$, mode $\hat{d}_\mathrm{B}$ equals $\hat{m}_\mathrm{B}$, and accordingly Eq.\,(\ref{Eq:HF_RoratingFrame}) reduces to $\hat{H}_F=\delta \hat{m}_\mathrm{D}^\dag \hat{m}_\mathrm{B}$. Therefore, the effect of the Floquet drive, which induces coupling between the center mode $\hat{d}_0$ and the two modes $\hat{d}_\pm$, is equivalent to inducing the coupling between the magnon dark and bright modes [Fig.\,\ref{fig4}(c)]. In this picture, the system consists of a cavity mode $\hat{c}$, a magnon bright mode $\hat{m}_\mathrm{B}$, and a magnon dark mode $\hat{m}_\mathrm{D}$. The interaction between the cavity and magnon bright modes is constantly on (with a coupling strength of $\sqrt{2}g$), while the interaction between the magnon dark and bright modes has a strength $\delta(\epsilon)$ which is controlled by the Floquet drive $\epsilon$. Therefore, this enables the on-demand bright-dark mode switching, allowing the isolation of the magnon modes in the two YIG spheres from the cavity as needed using an electronic signal. 

To demonstrate this capability, we performed a series of time-domain dynamical phase-field simulations based on coupled Maxwell-LLG equations \cite{Xu2024Mar,Zhuang2024Aug}, which eliminated the practical limitation of the finite magnon lifetimes. The validity of the numerical simulation is confirmed by comparison with the calculated spectra. The simulation results for a dual-sphere system with $2\Omega_0/2\pi=102$ MHz [which is different from Fig.\,\ref{fig4}(a)-(b)] and $\Delta=0$ are plotted in Fig.\,\ref{fig4}(d), and more simulations details can be found in the Supplemental Materials \cite{SM}. When a continuous-wave Floquet drive is applied (0-130 ns), the switching between the magnon dark and bright modes is constantly on, causing the Rabi-like oscillation between them after the dark mode is excited at $t=0$. The rapid oscillation between $\hat{c}$ and $\hat{m}_\mathrm{B}$ is due to the constantly on coupling between the two modes. When the system is in the magnon bright mode (130-160 ns), the application of a $\pi$ pulse (at 160 ns) switches the system to the magnon dark mode. The system stays in the dark mode, until another $\pi$ pulse (at 200 ns) switches it back to the magnon bright mode. Although the constant $\hat{c}$--$\hat{m}_\mathrm{B}$ coupling interferes with the magnon dark-bright mode switching, our numerical calculation shows that by properly selecting the coupling strength and Floquet drive, it is possible to completely suppress this effect \cite{SM}. 

Moreover, the response of the system to Ramsey pulse sequences (i.e., two $\pi/2$ pulses with a varying delay $\tau$) is also simulated. By adjusting the detuning ($\Omega_F-\Omega_0$) of the drive frequency in the Ramsey sequence, different interference fringes are observed [Fig.\,\ref{fig4}(e)]. As the detuning increases, the interference period becomes shorter, agreeing with the relation $T=2\pi/(\Omega_F-\Omega_0$). In conventional Ramsey interference experiments which typically involve two interacting modes, the system experiences a relaxation period between the two $\pi/2$ pulses. But in our three-mode system, there exists a constant interaction between the bright mode and the cavity mode. Consequently, the system stays for a shorter time in the $\hat{m}_\mathrm{B}$ mode than in the $\hat{m}_\mathrm{D}$ mode \cite{SM}, exhibiting different interference fringes [Fig.\,\ref{fig4}(e)].

%%%%%%%%%%%%%%%%%%%%%%%%%%%%%%%%%%%%%%%%%%%%%%%%%%%%%%%%%%%%%%%%
%%%%%%%%%%%%%%%%%%%%%%%%%%%%%%%%%%%%%%%%%%%%%%%%%%%%%%%%%%%%%%%%
%%%
%\section{conclusion}
%%%
%%%%%%%%%%%%%%%%%%%%%%%%%%%%%%%%%%%%%%%%%%%%%%%%%%%%%%%%%%%%%%%%
%%%%%%%%%%%%%%%%%%%%%%%%%%%%%%%%%%%%%%%%%%%%%%%%%%%%%%%%%%%%%%%%
To conclude, this work demonstrates a controlled approach that can isolate the magnon modes from the microwave cavity. Taking advantage of the collective mode in the effective giant spin ensemble formed by a dual-YIG sphere system, magnon dark mode can be switched on using a RF drive through the Floquet process, which has been recently introduced to hybrid magnonics but limited to hybrid modes only prior to this work. Such a mechanism can be used to isolate and retrieve magnon signals from the cavity, which is one of the key functionalities that have been missing for coherent information processing in cavity magnonics. Our approach can be scaled up to more complex systems with multiple magnonic resonances, pointing to a new direction for achieving large-scale, programmable integrated magnonic circuits.

\begin{acknowledgments}
X.Z. acknowledges support from NSF (2337713) and ONR Young Investigator Program (N00014-23-1-2144). L.J. acknowledges support from the ARO (W911NF-23-1-0077), ARO MURI (W911NF-21-1-0325), AFOSR MURI (FA9550-21-1-0209, FA9550-23-1-0338), NSF (ERC-1941583, OMA-2137642, OSI-2326767, CCF-2312755), Packard Foundation (2020-71479). The dynamical phase-field simulations in this work are supported by the US Department of Energy, Office of Science, Basic Energy Sciences, under Award Number DE-SC0020145 as part of the Computational Materials Sciences Program (Y.Z. and J.-M.H.). The dynamical phase-field simulations were performed using Bridges at the Pittsburgh Supercomputing Center through allocation TG-DMR180076 from the Advanced Cyberinfrastructure Coordination Ecosystem: Services \& Support (ACCESS) program, which is supported by NSF Grants No. 2138259, No. 2138286, No. 2138307, No. 2137603, and No. 2138296. J.M.J acknowledges support from AFOSR (FA9550-23-1-0254).
\end{acknowledgments}

\bibliography{MFCE_arxiv}% Produces the bibliography via BibTeX.

\end{document}

\title{Supplemental Materials for On-Demand Magnon Resonance Isolation in Cavity Magnonics}

\author{Amin Pishehvar}
\affiliation{ 
    Department of Electrical and Computer Engineering, Northeastern University, Boston, MA 02115, USA
}

\author{Zhaoyou Wang}
\affiliation{ 
    Pritzker School of Molecular Engineering, University of Chicago, Chicago, IL 60637, USA
}

\author{Yujie Zhu}
\affiliation{ 
    Department of Materials Science and Engineering, University of Wisconsin - Madison, Madison, WI 53706, USA
}

\author{Yu Jiang}
\affiliation{ 
    Department of Electrical and Computer Engineering, Northeastern University, Boston, MA 02115, USA
}

\author{Zixin Yan}
\affiliation{ 
    Department of Electrical and Computer Engineering, Northeastern University, Boston, MA 02115, USA
}

\author{Fangxin Li}
\affiliation{ 
    Pritzker School of Molecular Engineering, University of Chicago, Chicago, IL 60637, USA
}

\author{Josep M. Jornet}
\affiliation{ 
    Department of Electrical and Computer Engineering, Northeastern University, Boston, MA 02115, USA
}

\author{Jia-Mian Hu}
\affiliation{ 
    Department of Materials Science and Engineering, University of Wisconsin - Madison, Madison, WI 53706, USA
}

\author{Liang Jiang}
\affiliation{ 
    Pritzker School of Molecular Engineering, University of Chicago, Chicago, IL 60637, USA
}

\author{Xufeng Zhang}
\email{xu.zhang@northeastern.edu}
\affiliation{ 
    Department of Electrical and Computer Engineering, Northeastern University, Boston, MA 02115, USA
}
\affiliation{ 
    Department of Physics, Northeastern University, Boston, MA 02115, USA
}

\begin{minipage}[t]{\textwidth}
    % Create title on the same page
    \maketitle
    % Table of contents immediately follows
    \tableofcontents
\end{minipage}

\newpage

\section{Hamiltonian analysis}
The general Hamiltonian $\oph(t)$ for a two-mode Floquet cavity electromagnonics system is
\begin{equation}
    \begin{split}
        \oph(t) =& \oph_0 + \oph_F(t) \\
        \oph_0 =& \Delta_1 \opmd_1 \opm_1 + \Delta_2 \opmd_2 \opm_2 + g_1 (\opcd \opm_1 + \opc \opmd_1) + g_2 (\opcd \opm_2 + \opc \opmd_2) \\
        \oph_F(t) =& \varepsilon_1 \cos(\Omega_F t) \opmd_1 \opm_1 + \varepsilon_2 \cos(\Omega_F t + \phi) \opmd_2 \opm_2 .
    \end{split}
\end{equation}
We choose $\Delta_1 = -\Delta_2 \equiv \Delta \geq 0$ and $g_1 = g_2 \equiv g$ throughout the remaining texts.

%\section{Hamiltonian analysis}
We would like to explain certain features in the reflection spectrum from a Hamiltonian analysis, where we show a general derivation followed by a discussion of specific cases. Here we assume that the Floquet drive frequency matches the mode splitting and the Floquet drive amplitudes are relatively weak.

\subsection{Cavity reflection spectrum}
The cavity reflection (or transmission) spectrum is encoded in the frequency components of the Heisenberg picture operator $\opc(t)$, where $\opc$ is the cavity mode. This is true as long as the cavity linewidth is narrow enough and the reflection measurement is done at weak probe power. More concretely, consider a system with a lab frame Hamiltonian $\oph(t)$ which generates an evolution operator $\opu(t)$. Different frequency components $e^{-i\omega _n t}$ of $\opc(t) = \opud(t) \opc \opu(t)$ will appear as peaks at location $\omega_n$ in the reflection spectrum.

A simple example is a cavity with frequency modulation. The lab frame Hamiltonian is
\begin{equation}
    \oph(t) = (\omega_0 + \varepsilon \cos (\Omega t)) \opcd \opc,
\end{equation}
and the evolution operator is
\begin{equation}
    \opu(t) = \exp \left( -i \int_0^t \oph(t') dt' \right) =\exp \left( -i \opcd \opc \left( \omega_0 t + \frac{\varepsilon}{\Omega} \sin \Omega t \right) \right) .
\end{equation}
Therefore
\begin{equation}
    \opc(t) = \opud(t) \opc \opu(t) = \opc e^{-i \omega_0 t} \sum_{-\infty}^{\infty} J_n \left( \frac{\varepsilon}{\Omega} \right) e^{-i n \Omega t} ,
\end{equation}
which means that the reflection spectrum contains peaks at frequencies $\omega_0 + n \Omega$ and the peak heights are determined by amplitudes $J_n (\varepsilon/\Omega)$. To resolve different peaks, the cavity linewidth must be much smaller than $\Omega$.

For general systems, finding the exact evolution operator $\opu(t)$ will be difficult. However, with weak Floquet drive, we can calculate $\opc(t)$ perturbatively.

\subsection{General derivation}
Define the bright and dark magnon modes
\begin{equation}
    \opm_\pm = \frac{1}{\sqrt{2}} (\opm_1 \pm \opm_2) ,
\end{equation}
and the static Hamiltonian becomes
\begin{equation}
    \oph_0 = \Delta (\opmd_+ \opm_- +  \opmd_- \opm_+) + \sqrt{2} g (\opcd \opm_+ + \opc \opmd_+) .
\end{equation}
Notice that in $\oph_0$ only the bright magnon mode $\opm_+$ couples to the cavity mode $\opc$.

We can diagonalize $\oph_0$ as
\begin{equation}
    \oph_0 = \Omega_0 (\opdd_+ \opd_+ - \opdd_- \opd_-)
\end{equation}
where $\Omega_0 =  \sqrt{2g^2 + \Delta^2}$ and the hybridized modes are
\begin{equation}
    \begin{split}
        \opd_0 =& \frac{\sqrt{2} g}{\Omega_0} \opm_- - \frac{\Delta}{\Omega_0} \opc \\
        \opd_{\pm} =& \frac{1}{\sqrt{2}} \left( \left( \frac{\Delta}{\Omega_0} \opm_- + \frac{\sqrt{2} g}{\Omega_0} \opc \right) \pm \opm_+ \right) .
    \end{split}
\end{equation}
The inverse transformation is
\begin{equation}
    \begin{split}
        \opm_+ =& \frac{\opd_+ - \opd_-}{\sqrt{2}} \\
        \opm_- =& \frac{\Delta}{\Omega_0} \frac{\opd_+ + \opd_-}{\sqrt{2}} + \frac{\sqrt{2} g}{\Omega_0} \opd_0 \\
        \opc =&  \frac{\sqrt{2} g}{\Omega_0} \frac{\opd_+ + \opd_-}{\sqrt{2}} - \frac{\Delta}{\Omega_0} \opd_0 .
    \end{split}
\end{equation}

We can enter rotating frame by the evolution operator $\opu_0(t) = e^{-i \oph_0 t}$, which maps $\opd_+ \rightarrow e^{-i\Omega_0 t} \opd_+$, $\opd_- \rightarrow e^{i\Omega_0 t} \opd_-$ and $\opd_0 \rightarrow \opd_0$. In the rotating frame, the cavity mode operator becomes explicitly time-dependent
\begin{equation}
    \opc(t) = \opud_0(t) \opc \opu_0(t) = \frac{g}{\Omega_0} (e^{-i\Omega_0 t} \opd_+ + e^{i\Omega_0 t} \opd_-) - \frac{\Delta}{\Omega_0} \opd_0 .
\end{equation}
In the original frame, the Floquet drive Hamiltonian can be written as
\begin{equation}
    \begin{split}
        & \oph_F(t) \\
        =& \varepsilon_1 \cos(\Omega_F t) \opmd_1 \opm_1 + \varepsilon_2 \cos(\Omega_F t + \phi) \opmd_2 \opm_2 \\
        =& \frac{\varepsilon_1}{2} \cos(\Omega_F t) (\opmd_+ + \opmd_-) (\opm_+ + \opm_-) + \frac{\varepsilon_2}{2} \cos(\Omega_F t + \phi) (\opmd_+ - \opmd_-) (\opm_+ - \opm_-) \\
        =& \frac{\varepsilon_1 \cos(\Omega_F t)}{4 \Omega_0^2} ((\Delta + \Omega_0) \opdd_+ + (\Delta - \Omega_0) \opdd_- + 2g \opdd_0) ((\Delta + \Omega_0) \opd_+ + (\Delta - \Omega_0) \opd_- + 2g \opd_0) \\
        & + \frac{\varepsilon_2 \cos(\Omega_F t + \phi)}{4 \Omega_0^2} ((\Omega_0 - \Delta) \opdd_+ - (\Delta + \Omega_0) \opdd_- - 2g \opdd_0) ((\Omega_0 - \Delta) \opd_+ - (\Delta + \Omega_0) \opd_- - 2g \opd_0) .
    \end{split}
\end{equation}
We choose $\Omega_F = \Omega_0$ and assume $\varepsilon_1,\varepsilon_2 \ll \Omega_0$. In the rotating frame, only the interaction terms between $\opd_0$ and $\opd_\pm$ are static while all other terms such as $\opdd_+ \opd_+, \opdd_+ \opd_-$ are fast oscillating. Under RWA and keeping only the time-independent terms, the Floquet drive Hamiltonian in the rotating frame becomes
\begin{equation}
    \begin{split}
        \oph_F =& \frac{\varepsilon_1 g}{4 \Omega_0^2} \left[ ((\Delta + \Omega_0) \opdd_+ + (\Delta - \Omega_0) \opdd_-) \opd_0 + h.c. \right] \\
        & + \frac{\varepsilon_2 g}{4 \Omega_0^2} \left[ (-(\Omega_0 - \Delta) e^{-i \phi} \opdd_+ + (\Delta + \Omega_0) e^{i\phi} \opdd_-) \opd_0 + h.c. \right] \\
    =& \delta (\opbd \opd_0 + \opb \opdd_0) .
    \end{split}
\end{equation}
Here
\begin{equation}
    \delta = \frac{g \sqrt{(\varepsilon_1^2 + \varepsilon_2^2) (g^2 + \Delta^2) - 2 \varepsilon_1 \varepsilon_2 g^2 \cos \phi}}{2(2g^2 + \Delta^2)}
\end{equation}
and the mode operator $\opb = \alpha \opd_+ + \beta \opd_-$ with
\begin{equation}
    \begin{split}
        \alpha_ =& \frac{ \varepsilon_1 (\Delta + \Omega_0) - \varepsilon_2 (\Omega_0 - \Delta) e^{i \phi} }{2 \sqrt{(\varepsilon_1^2 + \varepsilon_2^2) (g^2 + \Delta^2) - 2 \varepsilon_1 \varepsilon_2 g^2 \cos \phi}} \\
        \beta =& \frac{ \varepsilon_1 (\Delta - \Omega_0) + \varepsilon_2(\Delta + \Omega_0) e^{-i\phi} }{2 \sqrt{(\varepsilon_1^2 + \varepsilon_2^2) (g^2 + \Delta^2) - 2 \varepsilon_1 \varepsilon_2 g^2 \cos \phi}} .
    \end{split}
\end{equation}
The mode orthogonal to both $\opb$ and $\opd_0$ is $\opb_0 = \beta^* \opd_+ - \alpha^* \opd_-$, and the inverse transformation is
\begin{equation}
    \begin{split}
        \opd_+ =& \alpha^* \opb + \beta \opb_0 \\
        \opd_- =& \beta^* \opb - \alpha \opb_0 .
    \end{split}
\end{equation}

We can diagonalize $\oph_F$ as
\begin{equation}
    \oph_F = \delta (\opbd_+ \opb_+ - \opbd_- \opb_-) ,
\end{equation}
where
\begin{equation}
    \opb_\pm = \frac{1}{\sqrt{2}} (\opb \pm \opd_0) .
\end{equation}
The inverse transformation is
\begin{equation}
    \begin{split}
        \opd_0 =& \frac{\opb_+ - \opb_-}{\sqrt{2}} \\
        \opd_+ =& \alpha^* \frac{\opb_+ + \opb_-}{\sqrt{2}} + \beta \opb_0 \\
        \opd_- =& \beta^* \frac{\opb_+ + \opb_-}{\sqrt{2}} - \alpha \opb_0 .
    \end{split}
\end{equation}
In the rotating frame of $\oph_F$ with $\opu(t)=e^{-i \oph_F t}$, the cavity mode becomes
\begin{equation}
    \begin{split}
        \opc(t) =& \opud(t) \left[ \frac{g}{\Omega_0} (e^{-i\Omega_0 t} \opd_+ + e^{i\Omega_0 t} \opd_-) - \frac{\Delta}{\Omega_0} \opd_0 \right] \opu(t) \\
        =& \frac{g}{\Omega_0} e^{-i\Omega_0 t} \left( \frac{\alpha^*}{\sqrt{2}} (e^{-i\delta t} \opb_+ + e^{i\delta t} \opb_-) + \beta \opb_0 \right) \\
        & + \frac{g}{\Omega_0} e^{i\Omega_0 t} \left( \frac{\beta^*}{\sqrt{2}} (e^{-i\delta t} \opb_+ + e^{i\delta t} \opb_-) - \alpha \opb_0 \right) \\
        & - \frac{\Delta}{\sqrt{2} \Omega_0} (e^{-i\delta t} \opb_+ - e^{i\delta t} \opb_-) .
    \end{split}
\end{equation}
By listing all frequencies components, the reflection spectrum in general contains 8 peaks at $\pm \Omega_0, \pm \Omega_0 \pm \delta, \pm \delta$. Here the top and bottom modes split into three peaks while the center mode only splits into two peaks. In specific cases which we will show below, the amplitudes at some frequencies in $\opc(t)$ may be 0, leading to certain peaks disappearing from the reflection spectrum.

\subsection{Special cases}

\subsubsection{$\Delta=0,\phi=0,\varepsilon_1=\varepsilon_2=\varepsilon$}
In this case, we have $\Omega_0=\sqrt{2}g$, $\alpha_\pm = \frac{1}{\sqrt{2}}$ and $\delta=0$. This leads to
\begin{equation}
    \opc(t) = \frac{1}{2} (\opc + \opm_+) e^{-i \sqrt{2}gt} + \frac{1}{2} (\opc - \opm_+) e^{i \sqrt{2}gt} .
\end{equation}
Therefore there are two peaks in the reflection spectrum at $\pm \sqrt{2} g$ with no splitting.

To understand what is going on, we can go back to the original Hamiltonian
\begin{equation}
    \oph = \sqrt{2} g (\opcd \opm_+ + \opc \opmd_+) + \varepsilon \cos(\Omega_F t) (\opmd_+ \opm_+ +  \opmd_- \opm_-) .
\end{equation}
Obviously the dark mode $\opm_-$ is decoupled from the relevant dynamics and the system is equivalent to the single magnon case in \cite{Xu2020Dec}. The reason that we don't have any splitting here is because we are considering $\Omega_F=\sqrt{2}g$ while the frequency difference between $\opd_+$ and $\opd_-$ is $2\sqrt{2} g$ and thus $\Omega_F=2\sqrt{2} g$ is required to see the AT splitting.

\subsubsection{$\Delta=0,\phi=\pi,\varepsilon_1=\varepsilon_2=\varepsilon$}
In this case, we have $\Omega_0=\sqrt{2}g$, $\alpha_\pm = \pm \frac{1}{\sqrt{2}}$ and $\delta=\varepsilon/2$, with $\opb_0=-\opc,\opb_+=\opm_1,\opb_-=\opm_2$. This leads to
\begin{equation}
    \opc(t) = \frac{1}{2\sqrt{2}} e^{-i \sqrt{2}g t} (e^{-i \frac{\varepsilon}{2} t} \opm_1 + e^{i \frac{\varepsilon}{2} t} \opm_2 + \sqrt{2} \opc) + \frac{1}{2\sqrt{2}} e^{i \sqrt{2}g t} (-e^{-i \frac{\varepsilon}{2} t} \opm_1 - e^{i \frac{\varepsilon}{2} t} \opm_2 + \sqrt{2} \opc) .
\end{equation}
Therefore both the top and bottom modes split into three peaks at $\pm \sqrt{2}g$ and $\pm \sqrt{2}g \pm \varepsilon/2$, while the center mode has no peaks.

The Floquet drive Hamiltonian in the rotating frame is
\begin{equation}
    \oph_F = \delta (\opmd_1 \opm_1 - \opmd_2 \opm_2) .
\end{equation}
The top mode $\opd_+$ is from the hybridization of $\opc$ and $\opm_+$. However due to the Floquet drive, in the rotating frame $\opm_+$ is no longer a normal mode since $\opm_1$ and $\opm_2$ are oscillating at different frequencies leading to 3 peaks in the spectrum.

\subsubsection{$\Delta\neq 0,\phi=\pi,\varepsilon_1=\varepsilon_2=\varepsilon$}
In this case, we have
\begin{equation}
    \Omega_0=\sqrt{2g^2 + \Delta^2}, \quad \alpha_\pm = \pm \frac{1}{\sqrt{2}}, \quad \delta = \frac{\varepsilon g}{\sqrt{2}\Omega_0}
\end{equation}
with
\begin{equation}
    \opb_0= -\frac{\Delta}{\Omega_0} \opm_- - \frac{\sqrt{2} g}{\Omega_0} \opc, \quad \opb_\pm = \frac{1}{\sqrt{2}} \left( \opm_+ \pm \left( \frac{\sqrt{2} g}{\Omega_0} \opm_- - \frac{\Delta}{\Omega_0} \opc \right) \right) .
\end{equation}
This leads to
\begin{equation}
    \begin{split}
        \opc(t) =& \frac{g}{2 \Omega_0} e^{-i\Omega_0 t} \left( e^{-i\delta t} \opb_+ + e^{i\delta t} \opb_- - \sqrt{2} \opb_0 \right) - \frac{g}{2 \Omega_0} e^{i\Omega_0 t} \left( e^{-i\delta t} \opb_+ + e^{i\delta t} \opb_- + \sqrt{2} \opb_0 \right) \\
        & - \frac{\Delta}{\sqrt{2} \Omega_0} (e^{-i\delta t} \opb_+ - e^{i\delta t} \opb_-) .
    \end{split}
\end{equation}
Decrease detuning $\Delta$ reduce the amplitudes $\Delta/\sqrt{2}\Omega_0$ of the center peaks at $\pm \delta$, making them harder to measure. The splitting $\delta$ indeed increases as $\Delta$ reduces. The top and bottom modes always have 3 peaks.  

\subsubsection{$\Delta\neq 0,\phi=0,\varepsilon_1=\varepsilon_2=\varepsilon$}
In this case, we have
\begin{equation}
    \Omega_0=\sqrt{2g^2 + \Delta^2}, \quad \alpha_\pm = \frac{1}{\sqrt{2}}, \quad \delta = \frac{\varepsilon g \Delta}{\sqrt{2}(2g^2 + \Delta^2)}
\end{equation}
with
\begin{equation}
    \opb_0= \opm_+, \quad \opb_\pm = \frac{\Delta \pm \sqrt{2}g}{\sqrt{2} \Omega_0} \opm_- - \frac{\Delta \mp \sqrt{2}g}{\sqrt{2} \Omega_0} \opc .
\end{equation}
This leads to
\begin{equation}
    \begin{split}
        \opc(t) =& \frac{g}{2 \Omega_0} e^{-i\Omega_0 t} \left( e^{-i\delta t} \opb_+ + e^{i\delta t} \opb_- + \sqrt{2} \opb_0 \right) + \frac{g}{2 \Omega_0} e^{i\Omega_0 t} \left( e^{-i\delta t} \opb_+ + e^{i\delta t} \opb_- - \sqrt{2} \opb_0 \right) \\
        & - \frac{\Delta}{\sqrt{2} \Omega_0} (e^{-i\delta t} \opb_+ - e^{i\delta t} \opb_-) .
    \end{split}
\end{equation}
The reflection spectrum is similar to the out of phase case with $\phi=\pi$. The only difference here is that the splitting $\delta$ is smaller than the out of phase case. More generally, we have
\begin{equation}
    \delta = \frac{\varepsilon g \sqrt{g^2 + \Delta^2 - g^2 \cos \phi}}{\sqrt{2}(2g^2 + \Delta^2)}
\end{equation}
which increases monotonically as $\phi$ goes from $0$ to $\pi$.

\subsubsection{$\phi=\pi,\varepsilon_1=\varepsilon,\varepsilon_2=0$}
In this case, we have
\begin{equation}
    \Omega_0=\sqrt{2g^2 + \Delta^2}, \quad \alpha_\pm = \frac{\Delta \pm \Omega_0}{2\sqrt{g^2 + \Delta^2}}, \quad \delta = \frac{\varepsilon g \sqrt{g^2 + \Delta^2}}{2(2g^2 + \Delta^2)}
\end{equation}
with
\begin{equation}
    \opb_0= \frac{1}{\sqrt{2(g^2 + \Delta^2)}} \left( \Delta \opm_+ - \Omega_0 \left( \frac{\Delta}{\Omega_0} \opm_- + \frac{\sqrt{2} g}{\Omega_0} \opc \right) \right) .
\end{equation}
The expressions for $\opb_\pm$ are long and thus omitted here. This leads to
\begin{equation}
    \begin{split}
        \opc(t) =& \frac{g}{2 \Omega_0 \sqrt{2(g^2 + \Delta^2)}} e^{-i\Omega_0 t} \left( (\Delta + \Omega_0) (e^{-i\delta t} \opb_+ + e^{i\delta t} \opb_-) + \sqrt{2} (\Delta - \Omega_0) \opb_0 \right) \\
        & + \frac{g}{2 \Omega_0 \sqrt{2(g^2 + \Delta^2)}} e^{i\Omega_0 t} \left( (\Delta - \Omega_0)  (e^{-i\delta t} \opb_+ + e^{i\delta t} \opb_-) - \sqrt{2} (\Delta + \Omega_0) \opb_0 \right) \\
        & - \frac{\Delta}{\sqrt{2} \Omega_0} (e^{-i\delta t} \opb_+ - e^{i\delta t} \opb_-) .
    \end{split}
\end{equation}

The spectrum is the same as the two-mode drive case where the top and bottom modes split into 3 peaks and the center mode disappears.

\subsection{Brief summary}
We first diagonalize the static part of the Hamiltonian
\begin{equation}
    \oph_0 = \Omega_0 (\opdd_+ \opd_+ - \opdd_- \opd_-) .
\end{equation}
In the rotating frame of $\oph_0$, the Floquet Hamiltonian becomes
\begin{equation}
    \oph_F = \delta \opdd_0 (\alpha \opd_+ + \beta \opd_-) + h.c. = \delta (\opbd_+ \opb_+ - \opbd_- \opb_-) ,
\end{equation}
where $|\alpha|^2 + |\beta|^2=1$. Notice that in the rotating frame $\opb_{\pm},\opb_0$ are the eigenmodes instead of $\opd_{\pm},\opd_0$. The inverse transformation is
\begin{equation}
    \begin{split}
        \opd_0 =& \frac{\opb_+ - \opb_-}{\sqrt{2}} \\
        \opd_+ =& \alpha^* \frac{\opb_+ + \opb_-}{\sqrt{2}} + \beta \opb_0 \\
        \opd_- =& \beta^* \frac{\opb_+ + \opb_-}{\sqrt{2}} - \alpha \opb_0 .
    \end{split}
\end{equation}

\section{Floquet scattering matrix}
\subsection{General formalism}
For a passive linear Floquet Hamiltonian $\oph (t)$ satisfying $\oph(t) = \oph(t+T)$ where $T = 2\pi/\Omega$, the Heisenberg equation is
\begin{equation}
    \begin{split}
        \ddt \bm{a} (t) =& i [\oph(t), \bm{a} (t)] - \frac{K}{2} \bm{a} (t) + \sqrt{K_e} \bm{a}_{\text{in}} (t) \\
        =& -i M(t) \bm{a} (t) - \frac{K}{2} \bm{a} (t) + \sqrt{K_e} \bm{a}_{\text{in}} (t) .
    \end{split}
\end{equation}
Here $\bm{a} = (\opa_1,...,\opa_n)$ are the $n$ modes in the system, $K_{e(i)} = \text{diag} (\kappa_{1,e(i)}, ..., \kappa_{n,e(i)})$ are the external (internal) loss rates of the modes and $K = K_e + K_i$ is the total loss rates.

Notice that $M(t) = M(t+T)$, we have the Fourier expansion
\begin{equation}
    M(t) = \sum_{k=-\infty}^{\infty} M_k e^{ik\Omega t} .
\end{equation}
Therefore the Heisenberg equation in the Fourier domain becomes an infinite dimensional coupled linear equations
\begin{equation}
    -i\omega \bm{a} (\omega) = -i \sum_{k=-\infty}^{\infty} M_k \bm{a} (\omega + k \Omega) - \frac{K}{2} \bm{a} (\omega) + \sqrt{K_e} \bm{a}_{\text{in}} (\omega) .
\end{equation}

We can truncate the Heisenberg equation at the $k$-th order, which gives
\begin{equation}
    G (\omega)
    \begin{pmatrix}
    \bm{a} (\omega - k\Omega) \\
    \vdots \\
    \bm{a} (\omega) \\
    \vdots \\
    \bm{a} (\omega + k\Omega) 
    \end{pmatrix}
    = (I_{2k+1} \otimes \sqrt{K_e})
    \begin{pmatrix}
    \bm{a}_{\text{in}} (\omega - k\Omega) \\
    \vdots \\
    \bm{a}_{\text{in}} (\omega) \\
    \vdots \\
    \bm{a}_{\text{in}} (\omega + k\Omega) 
    \end{pmatrix} ,
\end{equation}
where
\begin{equation}
    G (\omega) = 
    \begin{pmatrix}
    D_{\omega-k\Omega} & i M_1 & \cdots & & & \cdots & i M_{2k} \\
    i M_{-1} & D_{\omega-(k-1)\Omega} & i M_1 & \cdots & & & \vdots \\
    \vdots & \ddots & \ddots & \ddots & & & \vdots \\
    i M_{-k} & \cdots & i M_{-1} & D_\omega & i M_1 & \cdots & i M_k \\
    \vdots & & & \ddots & \ddots & \ddots & \vdots \\
    \vdots & & & & i M_{-1} & D_{\omega+(k-1)\Omega} & i M_1 \\
    i M_{-2k} & \cdots & & & \cdots & i M_{-1} & D_{\omega+k\Omega}
    \end{pmatrix}
\end{equation}
with
\begin{equation}
    D_\omega = -i (\omega I_n - M_0) + \frac{K}{2} .
\end{equation}

From the input-output relation, we have
\begin{equation}
    \begin{pmatrix}
    \bm{a}_{\text{out}} (\omega - k\Omega) \\
    \vdots \\
    \bm{a}_{\text{out}} (\omega) \\
    \vdots \\
    \bm{a}_{\text{out}} (\omega + k\Omega) 
    \end{pmatrix}
    = S(\omega)
    \begin{pmatrix}
    \bm{a}_{\text{in}} (\omega - k\Omega) \\
    \vdots \\
    \bm{a}_{\text{in}} (\omega) \\
    \vdots \\
    \bm{a}_{\text{in}} (\omega + k\Omega) 
    \end{pmatrix} ,
\end{equation}
where the scattering matrix is
\begin{equation}
    S(\omega) = (I_{2k+1} \otimes \sqrt{K_e}) G (\omega)^{-1} (I_{2k+1} \otimes \sqrt{K_e}) - I_{n(2k+1)} .
\end{equation}
The diagonal elements of the scattering matrix give the reflection coefficients.

\subsection{Floquet cavity electromagnonics}
For the two-mode Floquet cavity electromagnonics system, we have
\begin{equation}
    \begin{split}
        \oph(t) =& \oph_0 + \oph_F(t) \\
        \oph_0 =& \Delta_1 \opmd_1 \opm_1 + \Delta_2 \opmd_2 \opm_2 + g_1 (\opcd \opm_1 + \opc \opmd_1) + g_2 (\opcd \opm_2 + \opc \opmd_2) \\
        \oph_F(t) =& \varepsilon_1 \cos(\Omega_F t) \opmd_1 \opm_1 + \varepsilon_2 \cos(\Omega_F t + \phi) \opmd_2 \opm_2 .
    \end{split}
\end{equation}
Therefore $\bm{a} = (\opc,\opm_1,\opm_2)$ and
\begin{equation}
    M(t) = 
    \begin{pmatrix}
    0 & g_1 & g_2 \\
    g_1 & \Delta_1 + \varepsilon_1 \cos(\Omega_F t) & 0 \\
    g_2 & 0 & \Delta_2 + \varepsilon_2 \cos(\Omega_F t + \phi) 
    \end{pmatrix} .
\end{equation}
The non-zero Fourier components of $M(t)$ are
\begin{equation}
    M_0 = 
    \begin{pmatrix}
    0 & g_1 & g_2 \\
    g_1 & \Delta_1 & 0 \\
    g_2 & 0 & \Delta_2 
    \end{pmatrix}
    , \quad M_1 =
    \begin{pmatrix}
    0 & 0 & 0 \\
    0 & \frac{\varepsilon_1}{2} & 0 \\
    0 & 0 & \frac{\varepsilon_2}{2} e^{i\phi}
    \end{pmatrix}
    , \quad M_{-1} =
    \begin{pmatrix}
    0 & 0 & 0 \\
    0 & \frac{\varepsilon_1}{2} & 0 \\
    0 & 0 & \frac{\varepsilon_2}{2} e^{-i\phi}
    \end{pmatrix} .
\end{equation}
All other Fourier components $M_k$ are 0. In this case, the matrix $G(\omega)$ above becomes a tridiagonal block matrix.

\section{Time domain mode conversion}
We choose zero detunings and symmetric parameters to achieve the following Hamiltonian
\begin{equation}
	\begin{split}
		\oph(t) =& \sqrt{2} g (\opcd \opm_+ + \opc \opmd_+) + \varepsilon \cos(\Omega_F t + \phi) (\opmd_1 \opm_1 - \opmd_2 \opm_2)  \\
		=& \sqrt{2} g (\opcd \opm_+ + \opc \opmd_+) + \varepsilon \cos(\Omega_F t + \phi) (\opmd_+ \opm_- + \opm_+ \opmd_-) ,
	\end{split}
\end{equation}
where the bright and dark magnon modes are
\begin{equation}
\opm_\pm = \frac{1}{\sqrt{2}} (\opm_1 \pm \opm_2) .
\end{equation}

Since we are interested in the regime of $g \gg \varepsilon$, it is convenient to define
\begin{equation}
	\opa_{\pm} = \frac{1}{\sqrt{2}} (\opc \pm \opm_+) 
\end{equation}
and rewrite the Hamiltonian as
\begin{equation}
	\oph (t) = \sqrt{2} g (\opad_+ \opa_+ - \opad_- \opa_-) + \frac{\varepsilon}{\sqrt{2}} \cos(\Omega_F t + \phi) \left [ \opmd_- (\opa_+ - \opa_-) + \opm_- (\opad_+ - \opad_-) \right] .
\end{equation}

If $\Omega_F=0$, i.e., we just apply a static drive to shift the frequencies of $\opm_1,\opm_2$, it is obvious that we cannot achieve bright-dark mode conversion due to the large detuning between $\opm_-$ and $\opa_\pm$. In other words, a dark mode excitation will mostly stay in the dark mode and only with a small probability the excitation gets transfer to $\opa_\pm$.

We choose a Floquet drive with $\Omega_F = \sqrt{2} g$ to compensate for this large detuning and realize the bright-dark mode conversion. In the rotating frame of $\opa_\pm$, we have
\begin{equation}
	\begin{split}
		\oph (t) =& \frac{\varepsilon}{\sqrt{2}} \cos(\Omega_F t + \phi) \left [ \opmd_- (\opa_+ e^{-i\Omega_F t} - \opa_- e^{i\Omega_F t}) + \opm_- (\opad_+ e^{i\Omega_F t} - \opad_- e^{-i\Omega_F t}) \right] \\
		\approx & \frac{\varepsilon}{2\sqrt{2}} \left [ \opmd_- (\opa_+ e^{i\phi} - \opa_- e^{-i\phi}) + \opm_- (\opad_+ e^{-i\phi} - \opad_- e^{i\phi}) \right] \\
		=& \frac{\varepsilon}{2} \left [ \opmd_- \opB + \opm_- \opBd \right] ,
	\end{split}
\end{equation}
where
\begin{equation}
	\opB = i \sin (\phi) \opc + \cos (\phi) \opm_+ .
\end{equation}

Choosing $\phi=0$, we have
\begin{equation}
	\oph(t) = \frac{\varepsilon}{2} \left [ \opmd_- \opm_+ + \opm_- \opmd_+ \right] ,
\end{equation}
which realizes the conversion between magnon bright and dark mode. Choosing $\phi = \pi/2$, we have
\begin{equation}
	\oph (t) = i \frac{\varepsilon}{2} \left [ \opmd_- \opc - \opm_- \opcd \right] ,
\end{equation}
which realizes the conversion between microwave mode and the magnon dark mode. The conversion time is
\begin{equation}
	T_{\text{SWAP}} = \frac{\pi}{\varepsilon} .
\end{equation}

\begin{figure}[h]
	\centering
	\includegraphics[width=0.6\textwidth]{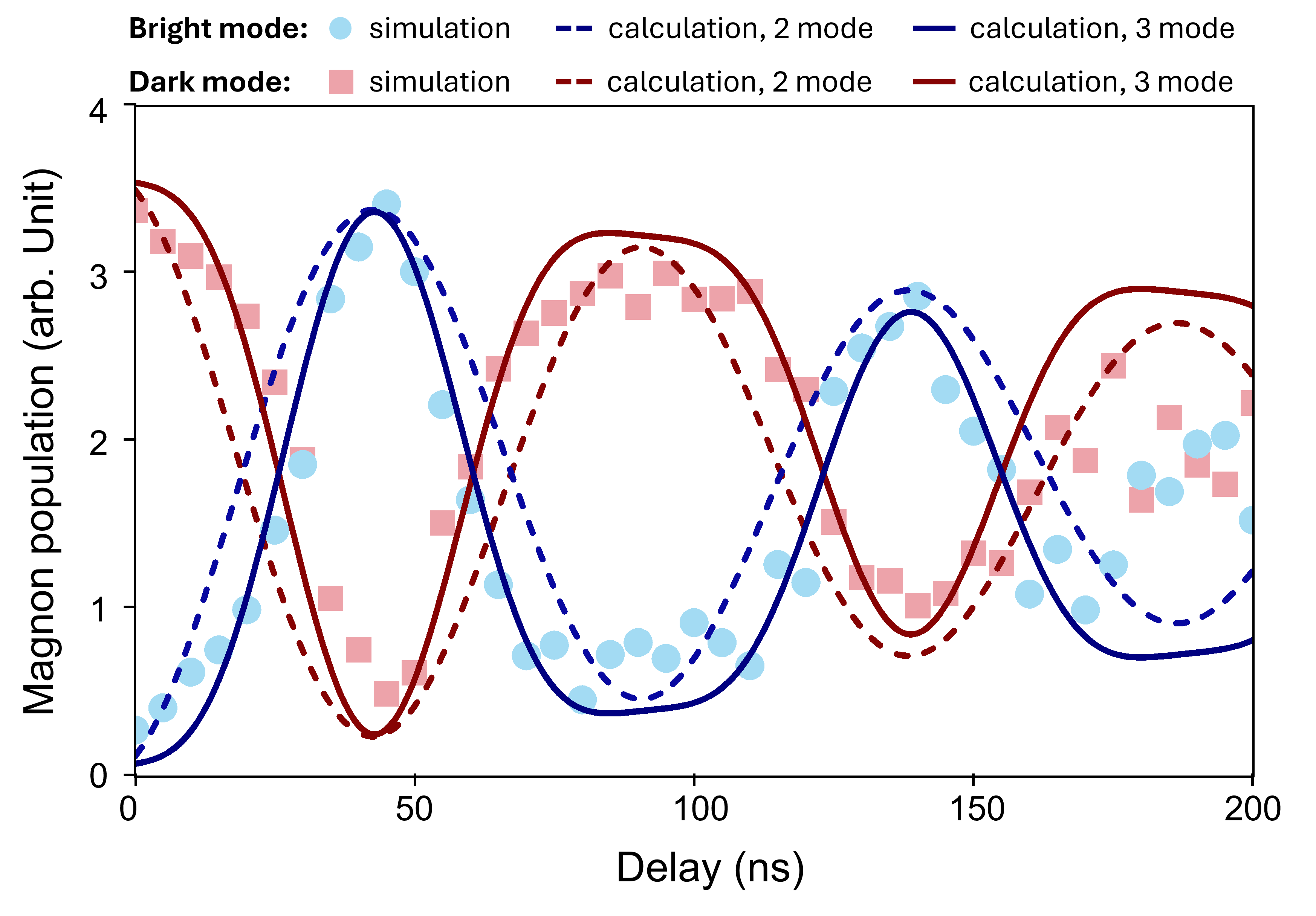}
	\caption{Comparison of the calculation results using the 2-mode Ramsey model (dashed lines) and our 3-mode model (solid lines) for bright (blue) and dark (red) modes, respectively. The simulation data are also plotted using blue circles (bright mode) and red squares (dark mode), respectively. The detuning of the driving field is $(\Omega_F-\Omega_0)/2\pi = 10$ MHz for both the calculation and simulation data.}
    \label{figSM:CMP}
\end{figure}

\subsection{Mode conversion with Ramsey sequence}
We can choose a non-zero detuning $\Omega_F = \sqrt{2}g + \Delta$ where $\Delta \ll g$. The Hamiltonian after RWA with $\phi=0$ is
\begin{equation}
	\oph = \Delta (\opcd \opm_B + \opc \opmd_B) + \frac{\varepsilon}{2} \left [ \opmd_D \opm_B + \opm_D \opmd_B \right] .
\end{equation}
Ramsey sequence (assume $\varepsilon\gg \Delta$):
\begin{enumerate}
    \item Initial state: $\ket{0}_C \ket{\sqrt{2}\alpha}_B \ket{0}_D$
    \item First $\pi/2$ swap: $\ket{0}_C \ket{\alpha}_B \ket{\alpha}_D$
    \item Delay $\tau$: $\ket{\alpha \sin(\tau \Delta)}_C \ket{\alpha \cos(\tau \Delta)}_B \ket{\alpha}_D$
    \item Second $\pi/2$ swap: $\ket{\alpha \sin(\tau \Delta)}_C \ket{-\frac{\alpha}{\sqrt{2}} (1-\cos(\tau \Delta))}_B \ket{\frac{\alpha}{\sqrt{2}} (1+\cos(\tau \Delta))}_D$
\end{enumerate}
Therefore the Ramsey signal in the 3-mode case, defined as the occupation of the dark mode, scales as $(1+\cos(\tau \Delta))^2$. This is different from the usual 2-mode Ramsey signal scales as $|1+\exp(i\tau \Delta)|^2$. From the comparison in Fig.\,\ref{figSM:CMP}, clearly the 3-mode model (solid lines) agrees well with our experimental data (squares and circles), while the 2-mode model (dashed lines) deviates quite significantly from the data. 

It is important to note that dissipation is omitted here for simplicity, as it has a negligible impact on the transition dynamics between dark and bright modes. Nevertheless, in the simulation results presented in Fig.\,4(e) and Fig.\,\ref{figSM:CMP}, the peak amplitudes of both the magnon dark and bright modes decrease over time with comparable decay rates across all three drive detuning scenarios. This decay in fact stems from numerical errors accumulated during the delay period between the two $\pi/2$ pulses. Since this effect modifies only the amplitudes instead of the periodicity of the interference pattern, the results in Fig.\,4(e) and Fig.\,\ref{figSM:CMP} still reliably capture the dynamics of transition between dark and bright modes (except that the amplitudes do not reflect the coherence time), which is evidenced by the close agreement between the simulation results and analytical calculations. Notably, our theoretical model incorporates a linearly decaying envelope, which is multiplied with the Ramsey signal described above, to enhance the visual alignment between these two sets of results.

\section{Experimental details}
Our experimental setup consists of a 3D cavity made of oxygen-free copper with interior dimensions of \(40 \times 9 \times 20\) mm\(^3\). A coaxial probe, positioned near the center of the cavity, probes the spectral response via reflection measurement (S\(_{11}\)) using a vector network analyzer (VNA). \textcolor{red}{The two YIG spheres are placed at the bottom wall of the cavity, symmetrically with respect to the cavity center, with a distance of 7.5 mm from the midpoint.} The frequencies of the magnon modes in the two YIG spheres can be independently tuned by utilizing the non-uniformity of the bias magnetic field provided by the permanent magnet. \textcolor{red}{The two MHz sources used for driving the coils are synchronized.}

\begin{figure}[h]
	\centering
	\includegraphics[width=0.6\textwidth]{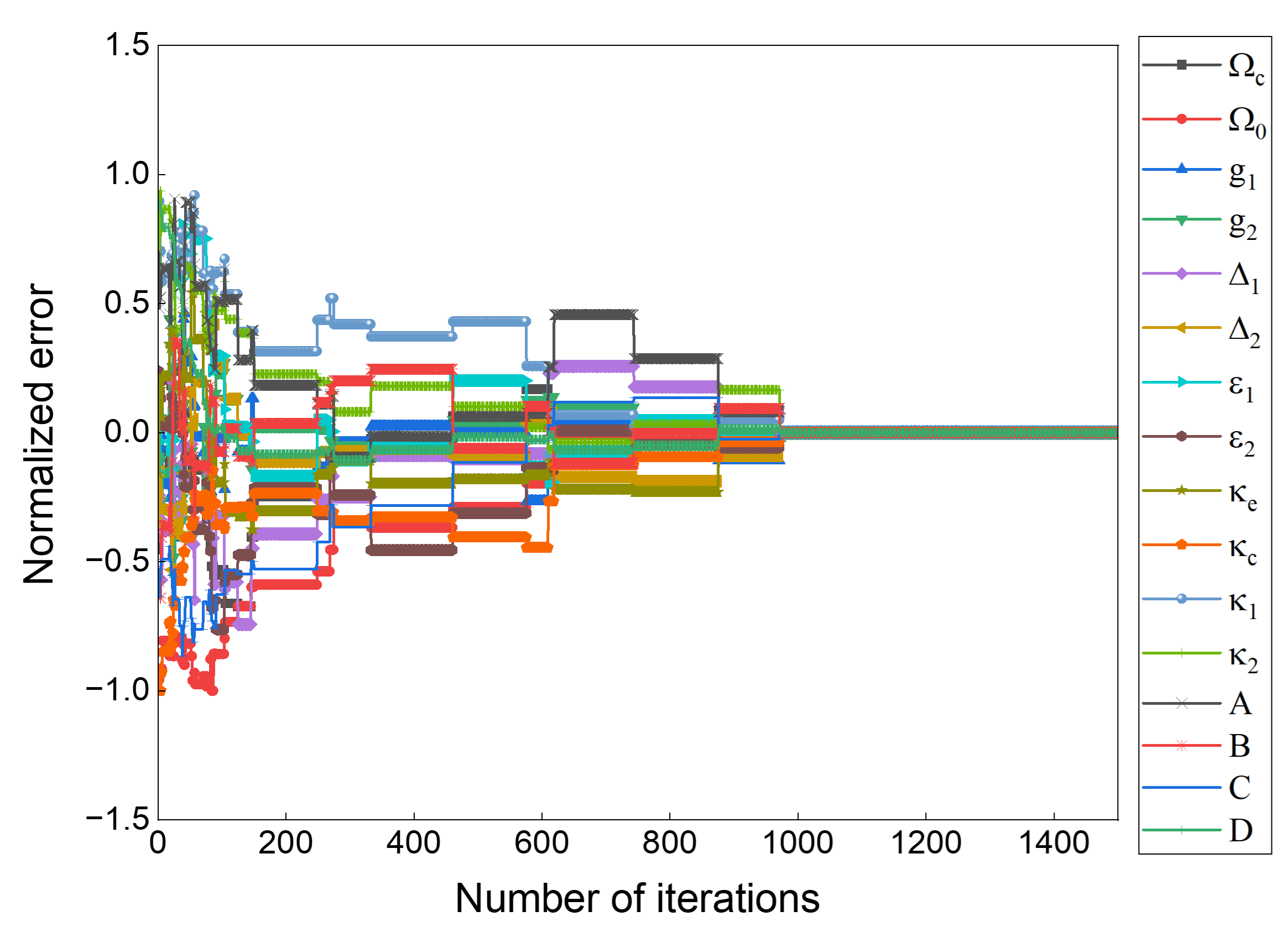}
	\caption{Convergence of the normalized errors for all the fitting variables.}
    \label{figSM:convergence}
\end{figure}

The parameters of the system configuration used in Fig.\, 2 of the main text are extracted using numerical fitting based on the above theoretical model with a polynomial background. The numerical fitting method is adopted from Ref.\cite{Zhang2019Dec_PRL}. A total of 16 variables are included in the fitting: the cavity frequency $\Omega_c$, intrinsic mode splitting $\Omega_0$, cavity external coupling rate $\kappa_c$, cavity dissipation rate $\kappa_c$, Magnon 1 dissipation rate $\kappa_1$, Magnon 2 dissipation rate $\kappa_2$, Magnon 1 detuning $\Delta_1$, Magnon 2 detuning $\Delta_2$, coupling strength of Magnon 1 with the cavity mode $g_1$, coupling strength of Magnon 2 with the cavity mode $g_2$, Magnon 1 driving amplitude $\varepsilon_1$, Magnon 2 driving amplitude $\varepsilon_2$, scaling factor of the background signal $A$, coefficient of the second order term in the polynomial background $B$, coefficient of the first order term in the polynomial background $C$, constant term of the polynomial background $D$. Here Magnon 1 (2) refers to the magnon mode (Kittel mode) in YIG sphere 1 (2). The convergence of all the parameters are shown in Fig.\,\ref{figSM:convergence}.

The extracted fitting parameters are listed below. Cavity resonance frequency: $\omega_c/2\pi=8.301$ GHz, external coupling rate of the cavity: $\kappa_e/2\pi = 0.7$ MHz, dissipation rate of the cavity resonance: $\kappa_c/2\pi = 6.1$ MHz, Magnon 1 dissipation rate: $\kappa_1/2\pi = 6.3$ MHz, Magnon 2 dissipation rate: $\kappa_2/2\pi = 3.1$ MHz, intrinsic mode splitting: $\Omega_0/2\pi=18.55$ MHz, coupling strength of Magnon 1 with the cavity mode: $g_1/2\pi=11.2$ MHz, coupling strength of Magnon 2 with the cavity mode: $g_2/2\pi=10.4$ MHz, detuning of Magnon 1 from cavity resonance: $\Delta_1/2\pi=11.5$ MHz, detuning of Magnon 2 from cavity resonance: $\Delta_2/2\pi=-11.5$ MHz, strength of the Floquet drive for Magnon 1: $\varepsilon_1/2\pi = 10.3$ MHz, strength of the Floquet drive for Magnon 2: $\varepsilon_1/2\pi = 6$ MHz.

\section{Time-domain simulations by dynamical phase-field modeling}

A GPU-accelerated dynamical phase-field model is employed to simulate the time-domain dynamics of the Floquet-drive-induced dark-bright mode transition shown in Fig.\,4 of the main text. The model enables simulating the bidirectionally coupled dynamics of magnons and cavity photons by numerically solving the coupled equation of motion for magnetization – the Landau-Lifshitz-Gilbert (LLG) equation – and the Maxwell’s equations. The model has previously been applied to simulate the dynamics of coherently coupled Kittel magnons and spoof surface plasmon polaritons (SSPPs) in a microstrip waveguide integrated with YIG resonators \cite{Xu2024Mar_slowwave}, as well as the dynamics of coherently coupled Kittel magnons and microwave photons in a cavity eletromagnonic system that contains one single YIG resonator placed in a 3D photon cavity \cite{Zhuang2024Aug_NPJCM}. Mathematical description of the model and its comprehensive validation have been detailed in \cite{Zhuang2024Aug_NPJCM}. It is noteworthy that our model also allows for simulating the fully coupled phonon-magnon-photon dynamics in multiphase systems, as shown in \cite{Zhuang2024Apr_PRAppl,Zhuang2023Jan_JPD,Zhuang2022Aug_NPJCM,Zhuang2021Oct_ACSAMI}. Details of the simulation set up for the present 3D cavity electromagnonic system that contains two YIG resonators as well as the materials parameters used for simulations are provided below.

\begin{figure}[b]
	\centering
	\includegraphics[width=0.8\textwidth]{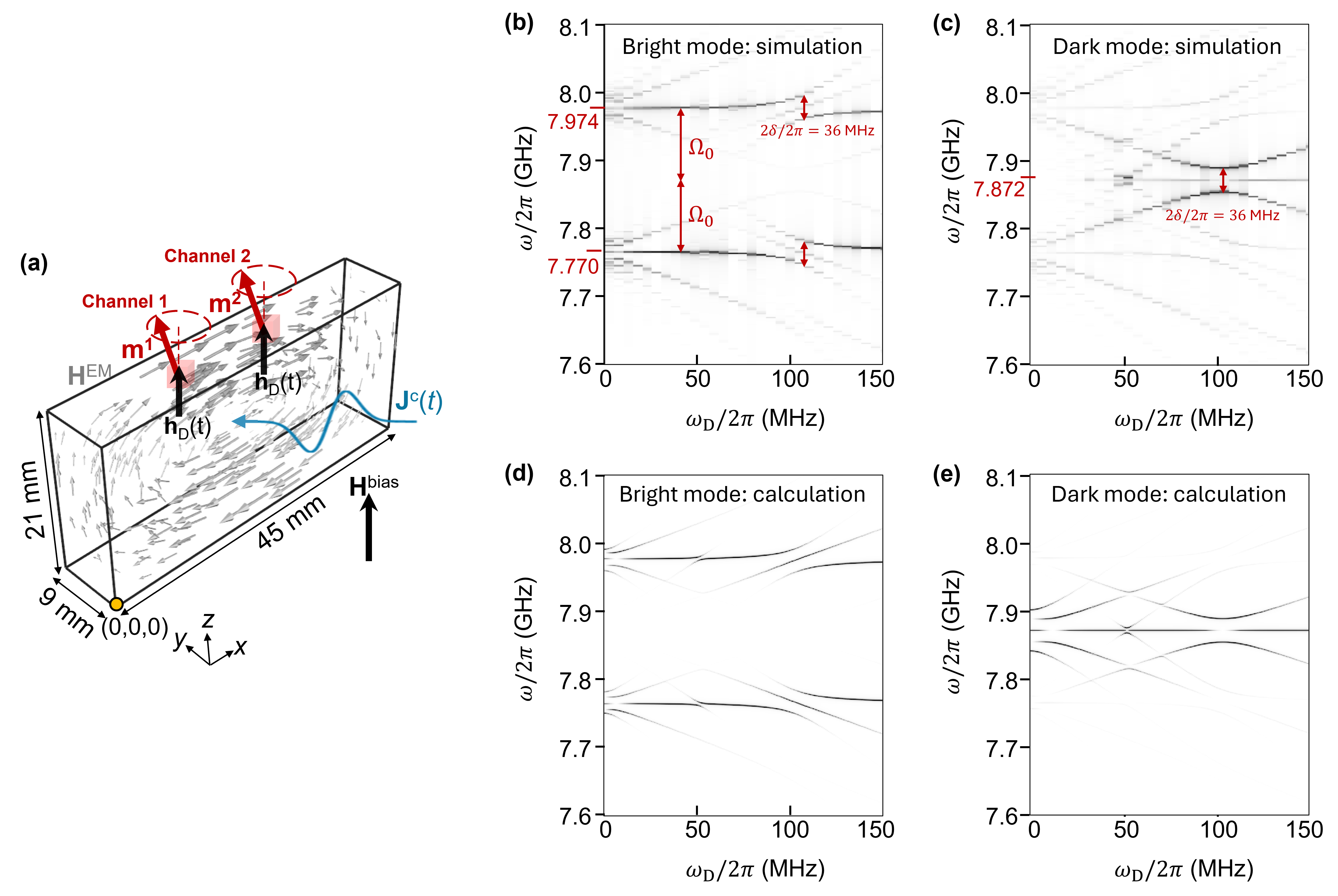}
	\caption{(a) Schematic of the 3D cavity electromagnonic system that integrates two cube-shaped YIG magnon resonators in a 3D microwave cavity. The vectors indicate the magnitude (proportional to arrow size) and direction (indicated by the arrowhead) of the local cavity magnetic field after the injection of a bipolar Gaussian current pulse at t=0 ns. Frequency spectra of (b) the magnon bright mode and (c) the magnon dark mode as a function of the Floquet drive frequency $\omega_D$. The Floquet drive is only applied to the YIG resonator at channel 2 with a fixed amplitude $h_{D,0}^2=2000$ A/m. An identical bias magnetic field of $H_0^\mathrm{bias}=231500$ A/m is applied to YIG resonators at both channels. (d) and (e) Calculated bright and dark mode spectra, respectively, using the same parameters as in (b)-(c) and based on the models in Sections I and II.}
    \label{figSM:time_domain_simulation}
\end{figure}

Figure \ref{figSM:time_domain_simulation} schematically shows the 3D cavity electromagonic system used in our simulations. The microwave photon cavity has a dimension of $45\times9\times21$ mm$^3$ which permits hosting the $TE_{101}$ cavity photon mode with a frequency of $\omega_c/2\pi=7.875$ GHz if assuming the dielectric permittivity $\varepsilon_r$ is 1 and isotropic, which is the case for the device used in our experiment. The two YIG resonators have the same dimension of $1\times1\times1$ mm$^3$. To excite the $TE_{101}$ cavity mode, a point current pulse $J^c(t)=J_0^c\frac{t}{2\sigma_0}\exp[-\frac{t}{2\sigma_0}]$ which only has the y-component, is applied at the position (22.5 mm, 1.5 mm, 10.5 mm) of the cavity. The Gaussian pulse duration parameter $\sigma_0$ is chosen to be 70 ps such that the frequency window covers the $\omega_c$. The amplitude $J_0^c$ is set to be $10^{15}$ A/m$^3$. With this value, the magnetic field component of the cavity photon (along x in the YIG resonators, see Fig. \ref{figSM:time_domain_simulation}) would be large enough to induce a sizable magnetization amplitude for the Kittel magnon mode (e.g.,$|\Delta{m}|=\sqrt{\Delta{m_x}^2+\Delta{m_y}^2}\approx10^{-3}$) but would not cause large-angle magnetization precision. 

The centers of the two YIG resonators locate at (x, y, z) = (20, 7.5, 10.5) mm (denoted as channel ‘1’) and (25, 7.5, 10.5) mm (denoted as channel ‘2’), respectively. Bias magnetic fields are applied along the $+z$ $([001])$ direction to stabilize the initial magnetization direction in the two YIG resonators along the $+z$ direction. The magnitudes of the bias magnetic fields applied to the YIG resonator at channel 1 and 2 are set as $H_{1}^\mathrm{bias}=H_{0}^\mathrm{bias}+\Delta{H^\mathrm{bias}}$ and $H_{2}^\mathrm{bias}=H_{0}^\mathrm{bias}-\Delta{H^\mathrm{bias}}$ , respectively, where $H_{0}^\mathrm{bias}=231500$ A/m. If there was no detuning ($\Delta{H^\mathrm{bias}}=0$), the FMR of both YIG resonators would be equal to the frequency of the cavity photon mode $\omega_c/{2\pi}=7.875$ GHz. The Floquet driving fields $h_D(t)$, which also only possess the z-component and take the form of $h_D(t)=h_{0}^{0}\sin(\omega_Dt)$ , are applied locally to the YIG spheres at channel 1 and channel 2. 

To ensure that (1) the YIG resonator only accommodates the Kittel mode magnon; and (2) reduce the EM wave propagation speed (such that larger timestep can be used to acceleration the time-domain simulations over a span of hundreds of ns, as in Fig. 4(d) of the main text, we increase the three diagonal components of the isotropic dielectric permittivity tensor $\varepsilon_r$ of both the photon cavity and the YIG resonators from 1 to $6.25\times10^{10}$. Detailed discussion and validation of this scaling strategy to simulate magnon-photon coupling in a 3D cavity electromagonic system are provided in \cite{Zhuang2024Aug_NPJCM}. In the specific system described above, this strategy allows us to downscale the size of the microwave cavity by $2.5\times10^5$ ($=\sqrt{\varepsilon_r}$) times from $45\times9\times21$ mm$^3$ to $180\times36\times84$ nm$^3$ because the EM wavelength is reduced by $\sqrt{\varepsilon_r}$ times. Likewise, the size of the YIG resonator is reduced proportionally from $1\times1\times1$ mm$^3$ to $4\times4\times4$ nm$^3$, which is much smaller than the exchange length of the YIG, $l_{ex}=\sqrt{A_{ex}/(0.5\mu_0M_S^2)}\approx16.3$ nm with an isotropic Heisenberg exchange coupling coefficient $A_{ex}$ of $3.26$ pJ/m and a saturation magnetization $M_S$ of $140$ kA/m \cite{Kamra2015Mar}. In this case, the exchange coupling ensures the local magnetization vectors (spins) in the YIG resonator precess as a macroscpin. In the present simulation set up, we use a 3D system of cuboid cells $N_x\Delta{x}\times N_y\Delta{y}\times N_z\Delta{z}$ (where $\Delta{x}$ = $\Delta{y}$ = $\Delta{z}$ = 2 nm, $N_x=90$, $N_y=18$, $N_z=42$) to discretize the 3D cavity electromagnonic system, where $2\times2\times2$ cells are used to represent the YIG resonator. A time step of $\Delta{t}=0.05$ ps is used to numerically solve the coupled LLG and Maxwell’s equations.

Figure \ref{figSM:time_domain_simulation} (b)-(c) show the simulated frequency spectra of the magnon bright mode and magnon dark mode under zero detuning ($\Delta{H^\mathrm{bias}}=0$) and the application of a Floquet drive field only to the YIG resonator at channel 2 (i.e., $h_D(t)=0$ for the YIG resonator at channel 1). The amplitude of the Floquet drive field $h_0^0$ is fixed at 2000 A/m yet the frequency of the Floquet drive $\omega_D/2\pi$ is varied linearly from 0 to 150 MHz with a step size of 5 MHz. Under such asymmetric Floquet drive condition, the temporal evolution of the magnon bright mode and dark mode can be extracted from the x or y component of the magnetization in the two YIG spheres. For instance, one can write $\langle \Delta m_x^1 \rangle = \Delta m_x^{\mathrm{bright}} + \Delta m_x^{\mathrm{dark}}$ and $\langle \Delta m_x^2 \rangle = \Delta m_x^{\mathrm{bright}} - \Delta m_x^{\mathrm{dark}}$, where $\Delta m_x^{\mathrm{bright}}(t)$ and $\Delta m_x^{\mathrm{dark}}(t)$ represent the time-domain evolution of the magnon bright mode and dark mode, respectively; $\langle \Delta m_x^1 \rangle$ is the volumetric average of the $\Delta m_x=m_x(t)-m_x(t=0)=m_x(t)$ in each cell of the YIG resonator at channel 1, and so forth for the $\langle \Delta m_x^2 \rangle$. In this regard, one can extract the temporal profile of the magnon bright and dark modes as, 
\begin{equation}
\Delta m_i^{\mathrm{bright}}(t) = \frac{\langle \Delta m_i^1 \rangle + \langle \Delta m_i^2 \rangle}{2}, \quad
\Delta m_i^{\mathrm{dark}}(t) = \frac{\langle \Delta m_i^1 \rangle - \langle \Delta m_i^2 \rangle}{2}, \quad i = x, y
\end{equation}

In this regard, the frequency spectra in Fig.\ref{figSM:time_domain_simulation} (b)-(c) are obtained by performing Fast Fourier Transform of the $\Delta m_x^{\mathrm{bright}}(t)$ and $\Delta m_x^{\mathrm{dark}}(t)$, respectively. Alternatively, the same spectral features can be obtained from $\Delta m_y^{\mathrm{bright}}(t)$ and $\Delta m_y^{\mathrm{dark}}(t)$.

Under zero detuning ($\Delta{H^\mathrm{bias}}=0$) and zero Floquet drive ($\omega_D=0$), the magnon bright mode has the same frequency as the cavity photon mode, and therefore hybridizes with the latter, resulting in the formation of two magnon polariton modes $\hat{d}_+$ and $\hat{d}_-$ at 7.974 GHz and 7.770 GHz, respectively. The frequency gap between the intrinsic cavity resonance frequency $\omega_c/2\pi=7.872$ GHz and the magnon polariton modes $\hat{d}_\pm$ is denoted as $\Omega_0$, as shown in Fig.\ref{figSM:time_domain_simulation} (b). The Floquet drive creates a series of sidebands of the $\hat{d}_\pm$ modes at frequencies $\omega_\pm+n\omega_\mathrm{F}$, where $n$ is the sideband order. When the frequency of the Floquet drive $\Omega_\mathrm{F}/2\pi=\Omega_0=102$ MHz, the first sideband ($n=\pm1$) of the dark mode $\hat{d}_0$ created by the Floquet drive resonantly interacts with the $\hat{d}_\pm$, leading to the so-called magnonic Autler-Townes splitting \cite{Xu2020Dec,Zhuang2024Aug_NPJCM} in the $\hat{d}_\pm$ mode with a frequency gap $2\delta/2\pi=36$ MHz. In contrast, when the Floquet drive is absent, the magnon dark mode $\hat{d}_0$ does not interact with the cavity photon mode and shows the same frequency as the cavity resonance at 7.872 GHz (i.e., $\omega_{\hat{d}_0}=\omega_c$). When the Floquet drive frequency $\Omega_\mathrm{F}=\Omega_0=102$ MHz, as shown in Fig.\ref{figSM:time_domain_simulation} (b), the the dark mode interact with the first lower sideband ($n=-1$) of the $\hat{d}_+$ mode and the first upper sideband ($n=1$) of the $\hat{d}_-$ mode, leading to a frequency splitting with a gap of $2\delta/2\pi=36$ MHz. This spectral feature is consistent with the experimental observation [c.f., Fig. 4(a) in the main text]. The frequency gap of 36 MHz indicates the rate of dynamical energy exchange between the magnon dark mode and the bright mode, corresponding to a time period $T_{ex}=2\pi/\Delta \omega_{AT}=28$ ns. Therefore, the $\pi$ pulse used for the time-sequence simulations in the main text has a duration of $T_\pi=T_{ex}/2=14$ ns. The spectra extracted from the dynamical phase-field simulation shown in Fig.\,\ref{figSM:time_domain_simulation} (b)-(c) exhibit good agreement with the theoretical spectra of the magnon bright and dark mode shown in Fig.\,\ref{figSM:time_domain_simulation} (d)-(e), which are calculated using the theoretical model described Section I and Section II and with the same parameters as used in Fig.\,\ref{figSM:time_domain_simulation}(b)-(c). Considering that our analytical model has already been shown to have excellent agreement with the experimental results (see Fig. 2 and Fig.4 (a)-(b) of the main text), such an agreement validates the accuracy of our numerical simulation approach. 

%\bibliography{MFCE_SM.bib} 

%apsrev4-2.bst 2019-01-14 (MD) hand-edited version of apsrev4-1.bst
%Control: key (0)
%Control: author (8) initials jnrlst
%Control: editor formatted (1) identically to author
%Control: production of article title (0) allowed
%Control: page (0) single
%Control: year (1) truncated
%Control: production of eprint (0) enabled
%